\documentclass[11pt]{article}
\usepackage{amssymb,amsmath}
\usepackage{doublespace}
\usepackage{cite} 
\usepackage{epsf}

\begin{document}
\begin{spacing}{1.0}
\title{ON THE GENERALIZED KRAMERS PROBLEM WITH OSCILLATORY MEMORY
FRICTION}
\author{Ram\'{o}n Reigada\\
Department of Chemistry and Biochemistry 0340\\
University of California San Diego\\
La Jolla, California 92093-0340\\
and\\
Departament de Qu\'{\i}mica F\'{\i}sica\\
Universitat de Barcelona\\
Av. Diagonal 647, E-08028 Barcelona, Spain\\
\and
Aldo H. Romero\\
Department of Chemistry and Biochemistry 0340\\
and Department of Physics\\
University of California, San Diego\\
La Jolla, California 92093-0340\\
\and
Katja Lindenberg\\
Department of Chemistry and Biochemistry 0340\\
and Institute for Nonlinear Science\\
University of California San Diego\\
La Jolla, California 92093-0340
\and
Jos\'{e} M. Sancho\\
Departament d'Estructura i Constituents de la Mat\`{e}ria\\
Universitat de Barcelona\\
Av. Diagonal 647, E-08028 Barcelona, Spain}

\date{\today }
\maketitle


\begin{abstract}
The time-dependent transmission coefficient for the Kramers problem
exhibits different behaviors in different parameter regimes.  In the
high friction regime it decays monotonically (``non-adiabatic"),
and in the low friction regime it decays in an oscillatory fashion
(``energy-diffusion-limited").  The generalized Kramers problem with an
exponential memory friction exhibits an additional oscillatory behavior in
the high friction regime (``caging").  In this paper we consider an
oscillatory memory kernel, which can be associated with a model in which
the reaction coordinate is linearly coupled to a nonreactive coordinate,
which is in turn coupled to a heat bath.  We recover the non-adiabatic and
energy-diffusion-limited behaviors of the transmission coefficient
in appropriate parameter regimes, and find that caging is not observed
with an oscillatory memory kernel.  Most
interestingly, we identify a new regime in which the time-dependent
transmission coefficient decays via a series of rather sharp steps
followed by plateaus (``stair-like").  We explain this regime and its
dependence on the various parameters of the system.

\end{abstract}

\section{Introduction}
\label{intro}

The classic Kramers formulation of reaction rates in solution~\cite{Kramers}
and its generalization to non-Markovian solvents~\cite{Grote}
has provided many theoretical challenges over the past six
decades~\cite{Hanggi,Melnikov2,Pollak,Tuckerrev}).
In this formulation
the reaction coordinate $x(t)$ is modeled as evolving in
a double-well potential $V(x)$ with a
barrier separating the reactant and product states.  The solvent effects
are modeled in terms of fluctuating and dissipative forces.
A full understanding
of the dependence of the rate coefficient $k$ on the
dissipation in the Markovian solvent limit (the ``turnover problem")
has only been achieved in the last few
years~\cite{Melnikov2,Melnikov1,Pollak1}. Comparably thorough
understanding in the case of a non-Markovian solvent is not yet available.
Understanding of the temperature dependence of $k$ is also far
from complete~\cite{Pollak2,Melnikov3,pendent1,pendent2}.
Clearly, there is yet a great deal to learn about this
classic problem.

In the past two decades, and most especially in the
past few years, attention has also been paid by a number of investigators
to the time-dependence of the rate coefficient, that is, the
way in which $k(t)$ approaches its asymptotic value
$k(\infty)$~\cite{pendent1,pendent2,Montg,Borgis,Tucker,Kohen}.
This time dependence directly mirrors the dynamics of the reaction
coordinate in the barrier region on the way toward capture by one well or
the other. Our focus is on this time dependence and the way that it is
influenced by the parameters of the system.

The generalized Kramers problem
is based on the dynamical equations for the reaction coordinate
\begin{equation}
\ddot{x} = -\int_0^t dt'~ \Gamma(t-t') ~ \dot{x}(t')
- \frac{dV_{eff}(x)}{dx} +F(t),
\label{generic1}
\end{equation}
where a dot denotes a time derivative, $V_{eff}(x)$ is an effective
potential related to $V(x)$ (cf. next section),
$\Gamma(t-t')$ is the
dissipative memory kernel (which we
will often simply call the {\em memory kernel}),
and $F(t)$ represents Gaussian
fluctuations that satisfy the fluctuation-dissipation
relation
\begin{equation}
\left< F(t)F(t')\right>=k_BT~\Gamma(t-t').
\label{cornoise}
\end{equation}
The brackets $\left<\cdots\right>$ denote an ensemble average,
$k_B$ is Boltzmann's constant, and $T$ is the temperature.
The fluctuations and dissipation account for the interaction of the
reaction coordinate with the surrounding medium.
The original Kramers problem dealt with a Markovian solvent, that is,
with instantaneous dissipation:
\begin{equation}
\label{original}
\Gamma(t)=2\gamma \delta(t).
\end{equation}
The parameter $\gamma$ is the dissipation parameter or damping parameter.
Generalizations to the non-Markovian problem have typically focused on
exponential memory kernels~\cite{Grote},
\begin{equation}
\Gamma(t)= \frac{\gamma}{\tau}\,e^{-t/\tau},
\label{expmem}
\end{equation}
and on oscillatory memory kernels~\cite{Grote},
\begin{equation}
\Gamma(t)= \Gamma(0)\,e^{-t/\tau}\left(\cos \Omega t +
\frac{1}{\Omega\tau}\sin \Omega t\right)
\label{oscillatory}
\end{equation}
(this memory kernel and the parameters in it will be discussed in detail
in the next section).  Another generalization, which we do not address in
our work, deals with Gaussian memory kernels~\cite{Grote,Tucker},
\begin{equation}
\Gamma(t)= \left(\frac{2}{\pi}\right)^{1/2}\frac{\gamma}{\tau}
e^{-t^2/2\tau^2}.
\label{gaussian}
\end{equation}
In all of these generalizations $\tau$ is a measure of the decay
time of the memory kernel or, equivalently, of the correlation time
of the fluctuations.

In subsequent sections we will provide a brief graphic review of the results
addressed in our previous work, which succinctly are as follows.
The time-dependent rate coefficient for the Markovian solvent at high
damping (that is, beyond the ``turnover" regime) decays monotonically
towards its equilibrium value~\cite{Kohen}. For the exponential memory at
high damping there are {\em two} distinct types of time dependences, the
``non-adiabatic", in which the rate coefficient decays monotonically
to its equilibrium value (as in the Markovian case), and the
``caging", in which the decay to equilibrium is oscillatory with
a frequency characteristic of an effective caging potential~\cite{Kohen}.
At very low damping (that is, below the ``turnover"),  the decay of
the rate coefficient to its equilibrium
value is again oscillatory, but now with a frequency characteristic of the
bistable potential.  This behavior is apparent for the Markovian
solvent~\cite{pendent1}
and also for the exponential memory~\cite{pendent2} in this
energy-diffusion-limited
regime.  We have shown that the theoretical predictions agree very well
with numerical simulations for all of these generic behaviors.

In this paper we complete our
analysis with a study of the oscillatory memory kernel.
In appropriate limits we recover the typical low-damping
behavior of the rate coefficient and also the high-damping
non-adiabatic monotonic behavior,
although caging, as we will see, can not be achieved with an
oscillatory memory.  Most
interesting, perhaps, is the appearance of a new time dependence
different from those previously observed or anticipated.  This new
time dependence is a
consequence of the new features of the memory kernel such as the fact that
it alternates between positive and negative values.  We will present
and explain this new behavior, and determine the parameter regimes where
it may be observed.

This paper is organized as follows. In Sec.~\ref{themodel} we introduce the
model Eq.~(\ref{generic1})
in detail.  In fact, we present two equivalent versions of the model.
One version invokes a solvent coordinate which is coupled to the reaction
coordinate and also coupled to a heat bath.  This
double presentation not only clarifies the physical origin of the
oscillatory memory kernel, but it also leads to more transparent
interpretation of the resulting time dependence of the rate coefficient.
In Sec.~\ref{simul} we describe our simulations and numerical procedures.
Section~\ref{results} presents a graphical summary of the
various time dependences obtained numerically
in earlier work and provides a context
for the presentation of the new behavior identified in the oscillatory
memory system.  In Sec.~\ref{approximations} we discuss analytic
approximations that serve as a backdrop for our analysis
and detailed explanation of the new behavior, which is presented in
Sec.~\ref{newreg}.  The results and conclusions are summarized in
Sec.~\ref{concl}.

\section{The Model}
\label{themodel}

We first present an alternative (two-variable) model that eventually
leads to
Eq.~(\ref{generic1}) with Eq.~(\ref{oscillatory}). The potential energy
of the two-variable model is
\begin{equation}
V(x,y) = V(x) + \frac{\omega^2}{2}y^2 + \frac{k}{2} (x-y)^2,
\label{potential}
\end{equation}
where the solvent is
explicitly represented by a harmonic coordinate $y$.
The reaction coordinate $x$ evolves in a bistable
potential that is taken to be of the familiar form
\begin{equation}
V(x)=\frac{V_0}{4}(x^2-1)^2.
\label{bistable}
\end{equation}
We take $V_0$ as the energy unit throughout this work and thus set it equal
to unity.  The reaction coordinate is coupled to the solvent coordinate
via a harmonic spring of force constant $k$.  The dynamical equations
for the coupled system, assuming that $y$ is coupled to a heat bath at
temperature $T$, are
\begin{eqnarray}
\ddot{x} = - \frac{dV(x)}{dx} +k ( y - x ),
\nonumber \\\nonumber\\
\ddot{y} = - \omega^2y + k ( x - y )
-\gamma \dot{y} + f(t).
\label{eqext}
\end{eqnarray}
Here $\gamma$ is the friction coefficient for the solvent coordinate
and $f(t)$ represents $\delta$-correlated Gaussian
fluctuations that satisfy the fluctuation-dissipation relation
\begin{equation}
\left< f(t)f(t')\right>=2\gamma k_BT\delta(t-t').
\label{fdr1}
\end{equation}
Throughout we take the barrier height to be large compared to the
temperature, $k_BT\ll 1/4$. We call Eq.~(\ref{eqext}) the ``extended"
representation of our system.

Although the extended model can readily be
integrated numerically, initial conditions are not arbitrary and
require careful consideration. The reduction of the model
(\ref{eqext}) to the generalized Kramers problem
(\ref{generic1}) with the fluctuation-dissipation relation (\ref{cornoise})
requires distributions for the initial solvent coordinate $y(0)$ and
velocity $\dot{y}(0)$ that satisfy certain conditions
(cf. Appendix~\ref{appa}, where these initial conditions are
presented in detail).

A number of theoretical approaches to this problem deal, instead, with the
completely equivalent ``contracted" or ``reduced" representation obtained
by explicitly integrating out the solvent coordinate $y$.  Among these is
the work of Grote and Hynes~\cite{Grote} and that of Kohen and Tannor
(KT) \cite{Kohen}.   The resulting equivalent single-variable problem is
shown in Appendix\ref{appa} to be given by Eq.~(\ref{generic1}) with
the effective potential
\begin{equation}
V_{eff}(x)=V(x) + \frac{1}{2} \frac{\omega^2 k}{\omega^2+k}\, x^2.
\label{efbistable}
\end{equation}
Depending on the relative values of parameters, the resulting memory
kernel can decay monotonically (``hyperbolic" case) or it can decay in an
oscillatory fashion (``trigonometric" case).  We are specifically interested
in the trigonometric case, which requires that
\begin{equation}
\left( \frac{\gamma}{2} \right) ^2 - \omega^2 - k < 0.
\label{und}
\end{equation}
The associated memory kernel is
\begin{equation}
\Gamma(t) = \frac{k^2}{\omega^2 +k}\,
e^{- \frac{\gamma}{2} t}\, \left( \frac{\gamma} {2 \Omega}
\sin \Omega t  + \cos \Omega t \right)
\label{trigfric}
\end{equation}
with the frequency
\begin{equation}
\Omega \equiv \sqrt{ \omega^2 + k - \left( \frac{\gamma}{2} \right) ^2}.
\label{omprima}
\end{equation}
We expect that the oscillatory
character of the memory
kernel may lead to new regimes of dynamical behavior that will become
evident in the time dependence of the rate coefficient.  We do not pursue
the hyperbolic case because we expect behavior similar
to that found earlier for the exponential memory
kernel and hence do not expect new behaviors in this case.

\begin{figure}[!htp]
\begin{center}
\hspace{3.in}
\epsfxsize = 4.in
\epsffile{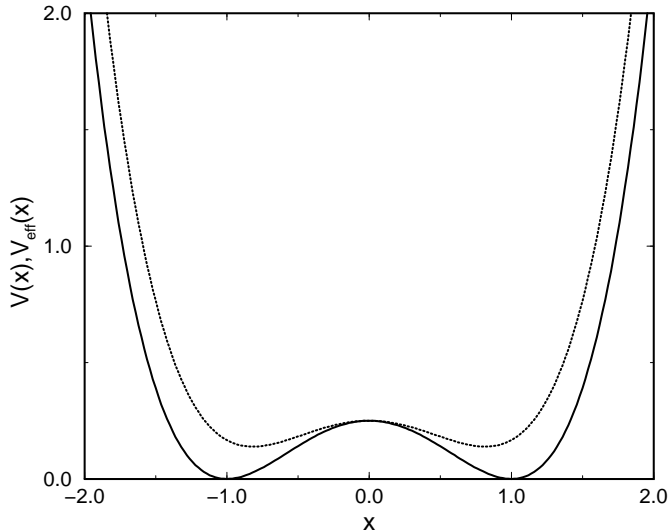}
\vspace{-0.5in}
\end{center}
\caption
{Solid line: original bistable potential $V(x)$ of Eq.~(\ref{bistable}).
Dotted line: effective potential $V_{eff}(x)$ of
Eq.~(\ref{efbistable}) for $\omega^2=0.5$ and $k=1$.}
\label{veff}
\end{figure}

From the explicit form (\ref{efbistable}) of the effective potential
we see that the additional quadratic term moves the
minima of the wells of the bistable potential $V(x)$ from $\pm 1$ to
\begin{equation}
x_{min}= \pm \sqrt{1-\frac{\omega^2 k}{\omega^2+k}}
\end{equation}
and diminishes the barrier from  $1/4$ to
\begin{equation}
\Delta V_{eff}^{\ddagger} = \frac{1}{4} - \frac{1}{2}
\left( \frac{\omega^2 k}{\omega^2+k} \right) +
\frac{1}{4} \left( \frac{\omega^2 k}{\omega^2+k} \right) ^2.
\end{equation}
Both effects can be seen in Fig.~\ref{veff}.
It is easily shown that the barrier disappears entirely
when $\omega^2 k \geq \omega^2 + k$, at which point the very
nature of the problem changes.  We thus constrain
our parameters $k$ and $\omega^2$ to ensure bistability:
\begin{equation}
\omega^2 k < \omega^2 + k.
\label{cons1}
\end{equation}

To summarize, then, the model to be considered in this paper is given
by Eqs.~(\ref{eqext}) with the potential (\ref{bistable}) and
the fluctuation-dissipation relation (\ref{fdr1}) (extended representation),
or, completely equivalently, by Eq.~(\ref{generic1}) with
the effective potential (\ref{efbistable}),
the memory kernel (\ref{trigfric}), and the fluctuation-dissipation
relation (\ref{cornoise}) (reduced representation).  Whichever
formulation is used, our parameters are constrained
by the inequality (\ref{cons1}), which ensures a bistable effective
potential, and by the inequality (\ref{und}), which ensures
an oscillatory memory kernel.  For some purposes the extended
representation provides a more convenient viewpoint, while for others the
reduced representation is more transparent.  In particular, we find the
extended representation more convenient for numerical simulations.

As a final point in this section it is important to note the altered
significance of parameters in the oscillatory memory kernel compared to the
Markovian or exponential models.  In the latter two cases
``high friction" and ``low friction" refer to the value of $\gamma$ since
this parameter directly measures the strength of the dissipative force that
extracts energy from the reaction coordinate into the bath.  Thus, in
the exponential memory case $\gamma$ is the value of $\Gamma(0)$ and also
of the integrated memory kernel.  In the case
under consideration here, however, $\gamma$  is a measure of the
dissipative force on the {\em solvent} coordinate and not directly on the
reaction coordinate.  Although $\gamma$ indirectly affects the loss of
energy of the reaction coordinate, the energy loss channel is now
principally determined by the coupling strength between the reaction
coordinate and the solvent coordinate.  Correspondingly,
now $\Gamma(0) =k^2/(\omega^2 +k)$.  The integrated memory kernel
is $\Gamma(0)\gamma/(\omega^2 +k)$, thus reflecting the overall
influence of $\gamma$.
However, it is the coupling constant $k$ that now essentially determines
whether we are in the ``high friction" or ``low friction" regime
(more details will be presented in Sec.~\ref{results}).

\section{Simulation Method: Initial Conditions and Other Details}
\label{simul}

The quantity of interest is the time-dependent rate coefficient $k(t)$
for an ensemble of particles evolving according to $x(t)$ in
Eq.~(\ref{generic1}) or Eq.~(\ref{eqext}). Numerically we find it more
convenient to work with the extended system (\ref{eqext}).
The coefficient $k(t)$ is the time-dependent mean rate of passage of the
particles across the barrier at $x=0$.  The usual focus on the deviation
of $k(t)$ from its equilibrium transition state
theory (TST) value~\cite{Hanggi,Melnikov2,pendent1} leads to the
expression
\begin{equation}
k(t)=\kappa(t) k^{TST}
\label{tst}
\end{equation}
where $k^{TST} = (\sqrt{2}/\pi)\exp(-1/4k_BT)$ is the
transition state theory rate that assumes that
particles never recross the barrier.  The transmission coefficient
$\kappa(t)$ is the correction to transition state theory
that includes both the temporal dynamics and the effects of
those particles that do recross the barrier.
We are interested in the dynamics
of the transmission coefficient $\kappa(t)$.

Numerically, one might try to calculate $k(t)$ directly
by starting all the particles
in one well and computing at each time how many of them
have crossed to the other well.  It would require an exceedingly long
calculation to gather statistically significant data in this manner,
since the reaction barrier is very much higher than the thermal
fluctuations.  The reactive flux formalism~\cite{Hanggi,Strauba,Straubb}
that relies on Eq.~(\ref{tst}) overcomes this difficulty since the
transmission coefficient can be calculated by dealing only with an ensemble
of particles whose initial position is above the barrier
[$x(0)=0$].  The slow process of ``getting there" is already included in
$k^{TST}$.  Half of the particles that start above the barrier have
a positive velocity
distributed according to the Boltzmann distribution in energy,
and the other half have the same distribution but with negative velocities.

Upon imposing the initial conditions on $y$ discussed in
Appendix~\ref{appa}, we
run quite a few iterations for the solvent coordinate evolution
in order to obtain even better thermalization.   Having achieved this, we
then integrate the fully coupled system (\ref{eqext}) with the following
distributions for the initial reaction coordinate position $x(0)\equiv
x_\circ$ and initial velocity $\dot{x}(0)\equiv v_{x\circ}$:
\begin{equation}
P(x_\circ)=\delta(x_\circ),
\label{x0}
\end{equation}
\begin{equation}
P(v_{x\circ}) = \frac{v_{x\circ}}{k_B T}\,
\exp\left(-\frac{v_{x\circ}^2}
{2 k_B T}\right).
\label{vx0}
\end{equation}

The numerical integration is carried out using the
second order Heun's algorithm~\cite{Gard,Toral94}. In all our runs
our ensemble consists of $N=10,000$ particles and we use
a very small time step ($\Delta t=0.001$). The transmission coefficient
is calculated from these
simulated data according to the relation~\cite{Strauba}
\begin{equation}
\kappa(t) = \frac{N_+ (t)}{N_+(0)} - \frac{N_- (t)}{N_-(0)},
\label{extract}
\end{equation}
where $N_+(t)$ and $N_-(t)$ are the particles that started with positive
velocities and negative velocities respectively and
at time $t$ are in or over the right hand well (i.e., the particles
for which $x(t)>0$).
Alternatively and completely equivalently (via
a simple symmetry argument) one can start all the $x$ particles with a
positive velocity and then
\begin{equation}
\kappa(t)=\kappa_+(t) -\kappa_-(t)
\label{extracto}
\end{equation}
where $\kappa_+(t)$ is the fraction of particles that are in or over the right
hand well at time $t$, and $\kappa_-(t)$ is the fraction in or over the left
hand well.  Furthermore, it is easily argued that~\cite{pendent1}
\begin{equation}
\kappa_-(t)=1-\kappa_+(t),
\label{symmetry}
\end{equation}
so it is sufficient to follow one or the other.

\section{Numerical Results}
\label{results}

\begin{figure}[htb]
\begin{center}
\hspace{3.in}
\epsfxsize = 5.in
\epsffile{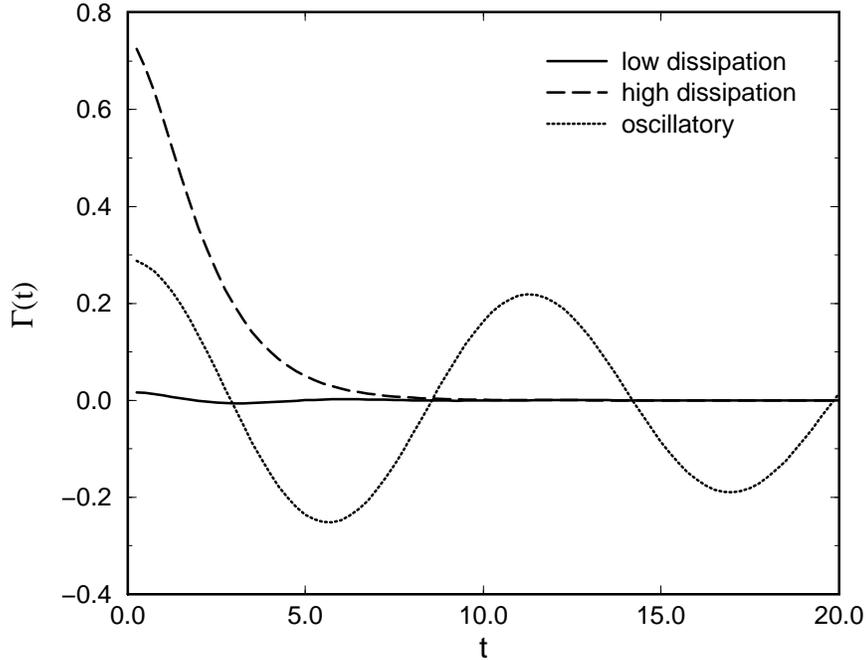}
\vspace{-0.5in}
\end{center}
\caption
{Memory kernel $\Gamma(t)$ vs $t$ for the different regimes studied
in Sec.~\ref{results}.
Solid curve: $\omega^2=1.0$, $k=0.14$, and $\gamma=0.667$.
Dashed curve: $\omega^2=0.01$, $k=0.75$, and $\gamma=1.74$.
Dotted curve: $\omega^2=0.01$, $k=0.3$, and $\gamma=0.05$.}
\label{diffric}
\end{figure}

The number of independent parameters in the generalized Kramers problem
with an oscillatory memory kernel is of course larger than for
Markovian or exponential frictions.  The values of $\kappa(t)$ and
$\kappa_{st}$ now in general depend on $k$, $\omega^2$, $\gamma$,
and $k_BT$.  Indeed, a systematic, even qualitative
study of the transmission coefficient as was done, for example,
in~\cite{pendent1,pendent2} would be quite complex.
We focus on a more modest goal, that is, to
capture qualitatively the different types of temporal behavior of
$\kappa(t)$ and the broad parameter regimes where each occurs.
These include the three regimes identified for the exponential memory,
namely, the energy-diffusion-limited, the non-adiabatic, and
the caging regimes, as well as possible new behaviors.

The oscillatory memory kernel can exhibit different appearances
depending on the parameter choices.  Figure~\ref{diffric}
exhibits three distinct ``generic" appearances, each roughly representative
of a distinct parameter regime.  Two of these mimic behaviors of the
exponential memory kernel and might be expected to lead to
transmission coefficients similar to those obtained earlier.
The third, the strongly oscillatory kernel, is new and might be
expected to lead to new behavior.  Let us consider each case in
turn, along with the resulting transmission coefficients.

The solid curve kernel in Fig.~\ref{diffric} mimics the exponential memory
kernel in the energy-diffusion-limited regime.  The kernel $\Gamma(t)$ is
small at all times.  In the case of the exponential memory kernel
(\ref{expmem}) this behavior was insured by choosing $\gamma$ to be small,
but in the oscillatory memory case, as noted earlier, the meaning of the
parameters is different.  Now a small value of $\Gamma(t)$ and, in
particular, a small value of $\Gamma(0)$ is insured if we choose small
values of $k^2/(\omega^2+k)$, that is, $k$ must be small and/or
$\omega^2$ must be large.  Note that the choices must still obey the
constraint (\ref{cons1}), but this is not a problem.  The value of $\gamma$
is not constrained by the low friction requirement, but it {\em is}
constrained by Eq.~(\ref{und}) if we want to insure that we are in the
oscillatory regime.  The value of $\gamma$ determines the oscillation
frequency of the memory kernel $\Gamma(t)$ but not its magnitude.

\begin{figure}[htb]
\begin{center}
\hspace{3.in}
\epsfxsize = 5.in
\epsffile{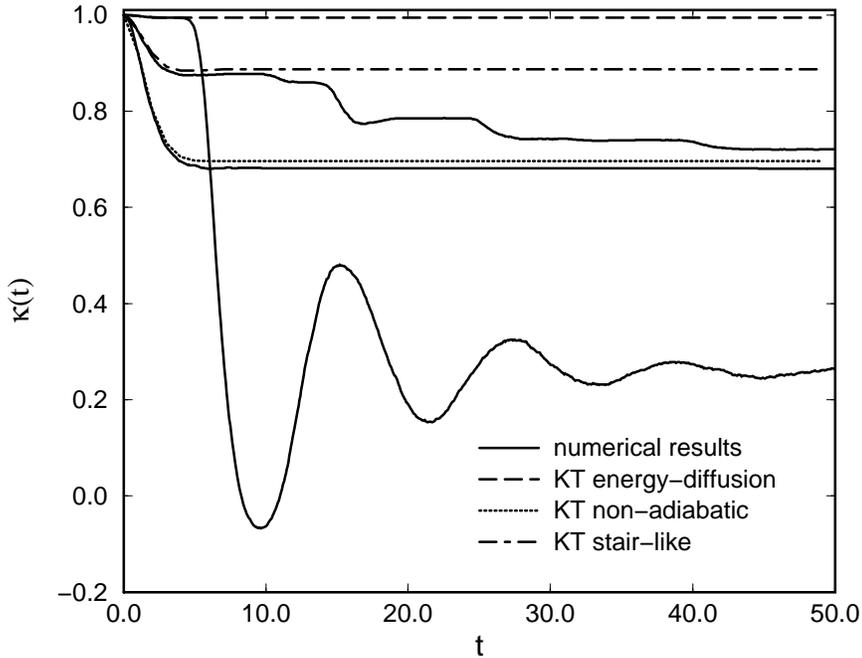}
\vspace{-0.5in}
\end{center}
\caption
{Solid curves: the three typical behaviors of the transmission coefficient
with oscillatory memory friction (numerical results).
Oscillatory curve: $\omega^2=1.0$, $k=0.14$, and $\gamma=0.667$
(energy-diffusion-limited regime).
Monotonic curve: $\omega^2=0.01$, $k=0.75$, and $\gamma=1.74$
(non-adiabatic regime).
The temperature for these two cases is $k_BT=0.025$.
Stepped curve: $\omega^2=0.01$, $k=0.3$, and $\gamma=0.05$ (new
``stair-like" regime).  The temperature is $k_BT=0.015$.
The dotted, dashed, and dot-dashed curves correspond to the KT theory
results for the same parameter values.}
\label{ktag}
\end{figure}

Typical values of the parameters that satisfy the conditions to produce
energy-diffusion-limited behavior while preserving the oscillatory
character of the memory kernel are
$\omega^2=1.0$, $k=0.14$, and $\gamma=0.667$.  These are the values
used to produce the solid curve in
Fig.~\ref{diffric}.  In this regime the
low dissipation causes the
dynamics of the system to be dominated by the slow
variation of the energy and consequently by
the repeated inertial recrossing of the barrier
before the particles are trapped in one well or the other.
As expected, we find
the typical energy-diffusion-limited behavior
for $\kappa(t)$ (oscillatory curve in
Fig.~\ref{ktag}) that consists of a very small initial decay to a plateau
up to a time beyond which $\kappa(t)$ decays in an oscillatory manner to
its equilibrium value.  As shown and discussed in our earlier
work~\cite{pendent1,pendent2}, the first decay is due to the few
low-energy particles that immediately change their initial direction due to
a thermal fluctuation and are trapped in the well opposite to the one
toward which they were initially moving. The oscillations are associated
with the essentially inertial successive recrossings of the higher-energy
particles.

All the arguments developed in the reduced system should have a
counterpart in the extended scheme. In the
energy-diffusion-limited case we have considered small $k$,
large $\omega^2$, and arbitrary $\gamma$ [subject to the
constraint (\ref{und})]. Large $\omega^2$ leads to a narrow
harmonic potential for $y$ [and consequently $y(t)$ remains small],
and small $k$ means weak coupling.  Since the coupling term $ky$
in Eqs.~(\ref{eqext}) provides the only energy loss channel
for the $x$ coordinate, large $\omega^2$ and small $k$ therefore lead
to low dissipation.

Let us now move on to the high dissipation regime.
The dashed curve in Fig.~\ref{diffric} mimics the exponential memory kernel
in the diffusion-limited regime.  The kernel $\Gamma(t)$ has a high initial
value and decays essentially monotonically (although we {\em are} in the
oscillatory regime).  In the case of the exponential memory kernel
(\ref{expmem}) this regime results when $\gamma$ is large.   For the
exponential memory kernel, the choice of the second parameter, $\tau$,
further determines two different regimes of behavior.  If the correlation
time $\tau$ is small, such that $\gamma/\tau<1$, the transmission
coefficient decays monotonically and one is said to be in the non-adiabatic
regime.  This is also the unique high-dissipation behavior associated
with the Markovian problem.  On the other hand, if $\tau$ is small, such
that $\gamma/\tau>1$, one is in the caging regime in which the behavior
of the transmission coefficient is oscillatory (but quite differently so
than in the low dissipation regime).  The
quasi-exponential dashed memory kernel in Fig.~\ref{diffric} corresponds to
the non-adiabatic behavior since the ratio of the initial value (about 0.8)
to the decay time of the kernel (about 4.0) is clearly smaller than unity.
In order to have
the oscillatory memory mimic the non-adiabatic exponential memory case as
in the figure
we require $\gamma$ to be {\em small} (since now $\gamma$ plays the
role that $1/\tau$ did before), and $\Omega$ to be small as well (to
minimize oscillatory effects).  A typical set of values that meets these
various conditions is $\omega^2=0.01$, $k=0.75$, and $\gamma=1.74$, which
leads to $\Omega=0.0557$.

The associated transmission coefficient for these parameters exhibits
the typical features of the
non-adiabatic regime, as shown by the monotonic curve in
Fig.~\ref{ktag}, namely, a smooth rather rapid decay
to the equilibrium value.
As in the case of an exponential memory, this decay in
the non-adiabatic regime looks Gaussian rather than exponential
at short times.

In the extended system small $\omega^2$ means that the potential in $y$
is very wide and for this reason $y(t)$ easily achieves large values.
Large $k$ represents strong coupling, and the combination of both
conditions leads to high dissipation for the $x$ coordinate.

As noted above, the other
regime found for an exponential memory kernel with high dissipation
is the
caging regime, which there occurs when both $\gamma$ and $\tau$ are
large, with $\gamma/\tau >1$.  Interestingly,
the constraints on the parameters and the shape
of the oscillatory friction kernel do not admit this regime.  This can be
understood from the following argument.
Caging is achieved when $\Gamma(t)$ is essentially constant over some
substantial time range so that
the friction integral in Eq.~(\ref{generic1}) over this time
can be approximated as a linear force on $x(t)$
and such that the resulting potential becomes monostable.
In our case, this resulting potential $V_r(x)$ would be
\begin{equation}
V_r(x)=\frac{1}{4}(x^2-1)^2 + \frac{1}{2} \frac{\omega^2 k}
{\omega^2+k} x^2 + \frac{1}{2} \frac{k^2}{\omega^2+k} x^2
= \frac{1}{4}(x^2-1)^2 + \frac{1}{2} k x^2.
\label{vres}
\end{equation}
From this expression it is easy to deduce that the potential $V_r(x)$ loses
its barrier when $k >1$.  The combination of the
condition that $\Gamma(t)$ behave roughly as a constant
for some time interval ($\gamma$ small) and that the resulting
potential lose its barrier during this time ($k>1$) would
lead to a caging regime with effective caging potential frequency
$\omega_{cag}=\sqrt{k-1}$. However, this combination of conditions can not
be satisfied with an oscillatory memory.
If we increase the value of $k$ above $1$, we also have to increase
$\gamma$ ($\Omega$ has to remain small to avoid
pronounced oscillations), but this in turn leads to the rapid exponential
decay of $\Gamma(t)$. It is thus not possible to achieve the
conditions for the caging regime
with trigonometric oscillatory friction.
The caging regime is easily captured in the hyperbolic case (cf.
Appendix~\ref{appa}),
since then $\Gamma(t)$ can take on a very high initial value that can be
sustained for a long time.

We have thus seen that the form of $\Gamma(t)$ and the constraints
on $k$, $\omega^2$,
and $\gamma$ determine which regimes typical
of exponential memories can also be captured with an oscillatory memory.
The requirements described so far have been met by either choosing
$\Gamma(0)$ to be small (low dissipation) or large
(high dissipation) while minimizing the amplitude of the oscillations.

{\em New behaviors} for the dynamics of $\kappa(t)$ may appear in
parameter regimes that emphasize the oscillatory behavior of $\Gamma(t)$.
To provide such emphasis we minimize the
damping effects of the exponential part by choosing
$\gamma$ to be small. Further choosing
a very small value for $\omega^2$ and a medium value for $k$
(large enough to get high amplitudes of oscillation of $\Gamma(t)$,
but limited so that the frequency $\Omega$ is not too high)
we obtain the oscillatory friction shown as the dotted kernel in
Fig.~\ref{diffric}. Since $\gamma$ and $\omega^2$ are very
small, the frequency of the oscillations is
\begin{equation}
\Omega \approx \sqrt{k}.
\label{wfr}
\end{equation}
In this regime at low temperatures an entirely new temporal behavior
emerges for $\kappa(t)$.
We generically call this the ``stair-like" regime.  It is shown for a
typical set of parameter values in Fig.~\ref{ktag}
(corresponding to those of the dotted curve in
Fig.~\ref{diffric}).
The main feature of this new behavior
is that $\kappa(t)$ exhibits a ``stair" shape,
namely, it decays via a series of steps followed by plateaus.
The explanation of this behavior, including the period of the steps,
the dependences on the parameters $\gamma$, $\omega^2$, $k$, and $k_BT$,
and the connections with the other regimes are presented in
detail in Sec.~\ref{newreg}.

\section{Approximations}
\label{approximations}

In this section we lay the groundwork
for the arguments invoked in the next section, where we discuss
the stair-like regime in detail.
Our explanations are semi-quantitative, that is, we do not
develop a theory that reproduces the stair-like curve in the figure in all
its details.  We mention this because in fact such theories {\em are}
available for the two other curves~\cite{Borgis,Kohen,pendent1,pendent2}.
In the high dissipation regime the KT
theory predicts the monotonic decay in the non-adiabatic
regime, and this prediction, shown in Fig.~\ref{ktag},
is seen to be quantitatively very good (see also~\cite{pendent1,pendent2}).
In the low dissipation regime we have shown that our theory also leads to
very good quantitative agreement with numerical results for the entire time
evolution of $\kappa(t)$~\cite{pendent1,pendent2}.  We have not
derived such a detailed formula for the stair-like regime, but we
nevertheless have been able to gain considerable understanding of this
behavior, and this is what we shall present.

Our insights turn out to be most complete if we invoke both
representations of the oscillatory problem, the extended as well as
the reduced. Furthermore, an understanding of the early time
dependence of $\kappa(t)$ in both of these representations turns out to be
very helpful, even if the approximations that are invoked are not valid
over the entire time regime -- the breakdown of approximations can also
yield useful insights.  We thus first turn to the early time behavior.

Consider first the reduced representation.  KT theory
focuses on the way in
which particles subject to dynamics of the generic form (\ref{generic1})
with the potential approximated by a parabolic barrier
diffuse to one side or the other of the barrier.
Since their analysis is restricted to a parabolic barrier (rather
than a bistable potential), the theory
is appropriate only for high dissipation, that is, when
the reaction coordinate never recrosses the barrier once it has left the
barrier region.  In other cases KT theory may (and indeed does)
capture only the initial decay, typically up to the first plateau value
of $\kappa(t)$, but it does not capture the asymptotic values
$\kappa_{st}$.
This is seen in Fig.~\ref{ktag}, where the KT theory predictions are shown
for each of the generic transmission coefficients.  KT theory works
very well for all times for the non-adiabatic high dissipation curve, and
captures the initial decay in the energy-diffusion-limited (dashed curve)
and stair-like (dot-dashed curve) cases.  These initial agreements
are fairly typical for all parameter values.

\begin{figure}[htb]
\begin{center}
\leavevmode
\epsfxsize = 3.0in
\epsffile{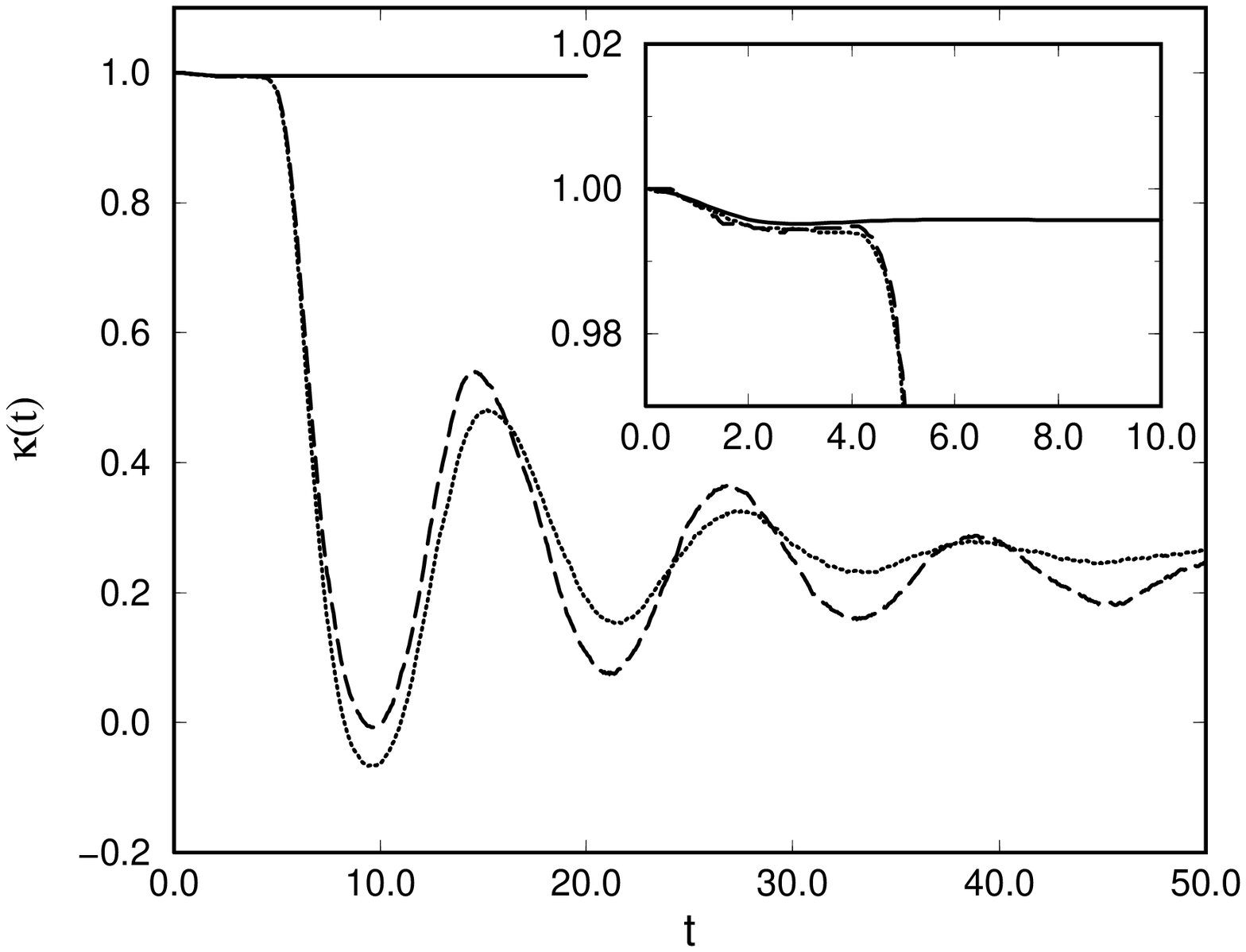}
\leavevmode
\epsfxsize = 3.0in
\epsffile{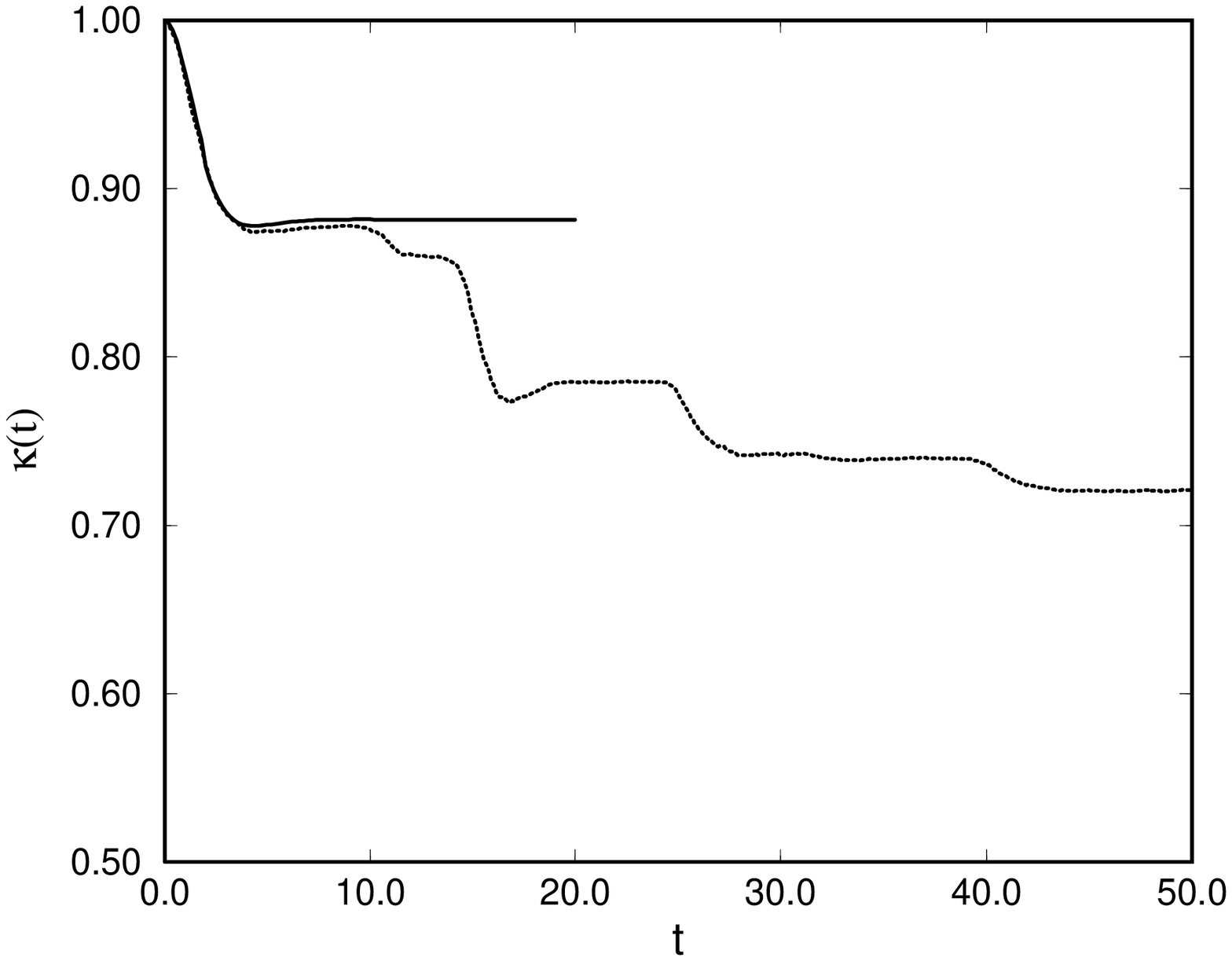}
\vspace{-0.2in}
\end{center}
\caption
{Left panel: The dotted and dashed curves are numerical simulations of
the transmission
coefficient $\kappa(t)$ vs $t$ in the energy-diffusion-limited regime.
In both cases $\omega^2=1.0$, $k=0.14$ and $k_BT=0.025$.  Dotted curve:
$\gamma=0.667$ (same value as in Fig.~\ref{ktag}); dashed curve:
$\gamma=0.05$.
The solid line, which reproduces both early time behaviors very
closely (see inset), is obtained from our early-time approximation for
the extended system, Eq.~(\ref{detext}).
Right panel: the dotted curve is our simulation in the stair-like regime
(cf. Fig.~\ref{ktag}).  The solid curve is the early-time approximation.
}
\label{agexta}
\end{figure}

Consider now the extended representation (\ref{eqext}).  We introduce
an even simpler approximation in this representation
that also captures the early time behavior when the dissipation
of energy in the $x$-coordinate is slow
and that facilitates our analysis of the stair-like regime.
The approximation is based on three main assumptions, all appropriate
only at short times. One is
akin to the argument we used earlier in the energy-diffusion-limited
problem, namely, that the main influence of the temperature
arises from the initial thermal
distributions.  Thus, as long as the initial distributions are chosen
correctly, that is, according to Eqs.~(\ref{y0}), (\ref{vy0}),
(\ref{x0}), and
(\ref{vx0}), the thermal effects in the form of the explicit random force
acting on the solvent coordinate can be omitted from the dynamical
equations.  The second is the omission of the dissipation term, i.e., we
set $\gamma$ to zero.  Note that this is the dissipative force on the
solvent coordinate; the principal initial dissipative channel for the
reaction coordinate $x(t)$ is its coupling to the $y$ coordinate via $k$,
and this is retained.  The third is to use a parabolic
barrier to approximate the potential.  With these assumptions the initial
decay of the transmission coefficient is due to the low-energy particles
(i.e. those barely above the barrier) that are pulled by the $y$ coordinate
in a direction opposite to the one indicated by their initial velocity, as
described by the simplified deterministic coupled linear equations
\begin{align}
\ddot{x}(t)&= (1-k) x + k y
\notag \\ \notag\\
\ddot{y}(t)&= - (\omega^2 + k) y + k x .
\label{detext}
\end{align}
With the initial distributions (\ref{y0}), (\ref{vy0}), (\ref{x0}) and
(\ref{vx0}) we can then use the form Eq.~(\ref{extracto}) for the
transmission coefficient to write
\begin{equation}
\kappa(t) = \int_{-\infty}^{\infty} dv_{y\circ}
\int_{-\infty}^{\infty} dy_\circ \int_0^{\infty} dv_{x\circ}\,
P(v_{x\circ})\,P(y_\circ)\,P(v_{y\circ}){\rm sgn}[x(t;v_{x\circ},y_\circ,
v_{y\circ})]
\label{integral}
\end{equation}
where ${\rm sgn}[x]$ is the
sign function, that is, ${\rm sgn}[x]=+1$ if
$x>0$ and $-1$ if $x<0$, and
$x(t;v_{x\circ},y_\circ,v_{y\circ})$ is the solution of
Eqs.~(\ref{detext}) with initial conditions
$v_{x\circ}$, $y_\circ$, $v_{y\circ}$, and $x_\circ=0$.

The left panel in
Fig.~\ref{agexta} shows two simulations in the energy-diffusion-limited
regime along with the results of
(\ref{integral}) with (\ref{detext}).  The early time
agreement in this regime is clearly excellent, as seen
in the detail inset.  Note that both simulations exhibit the same
early-time behavior, even though the value of $\gamma$ is very different for
the two cases (and not particularly ``small" in one of the two cases).
Clearly, in this regime the values of $k$ and $\omega^2$
determine the early time behavior of the transmission coefficient.

More importantly for our purposes here, the right panel of Fig.~\ref{agexta}
shows similar early-time agreement between the stair-like numerical
results and the approximation.  The agreement extends through the first plateau.
Note that the approximation captures the (slightly) non-monotonic behavior of the
simulation results.  The agreement between the two curves provides
the basis for our analysis of the stair-like regime in the next section.

\section{The Stair-like Regime}
\label{newreg}

We saw in Sec.~\ref{results} that the stair-like regime is achieved
when $\gamma$ and $\omega^2$ are small and the temperature is low.
The main feature
of this new behavior is that the transmission coefficient shows progressive
decays connected by plateaus. The length of the plateaus
(determined by a time period that we call $T_{\kappa}$),
the depths of the decays,
and all the characteristics that define this regime depend on the
values of the
parameters, principally $k$. Our understanding of this regime is based
on argumentation that relies mainly on the extended system, although
some of the arguments can easily be translated to the
language of the reduced scheme.

\subsection{Trajectories}
\label{traject}

\begin{figure}[htb]
\begin{center}
\leavevmode
\epsfxsize = 3.0in
\epsffile{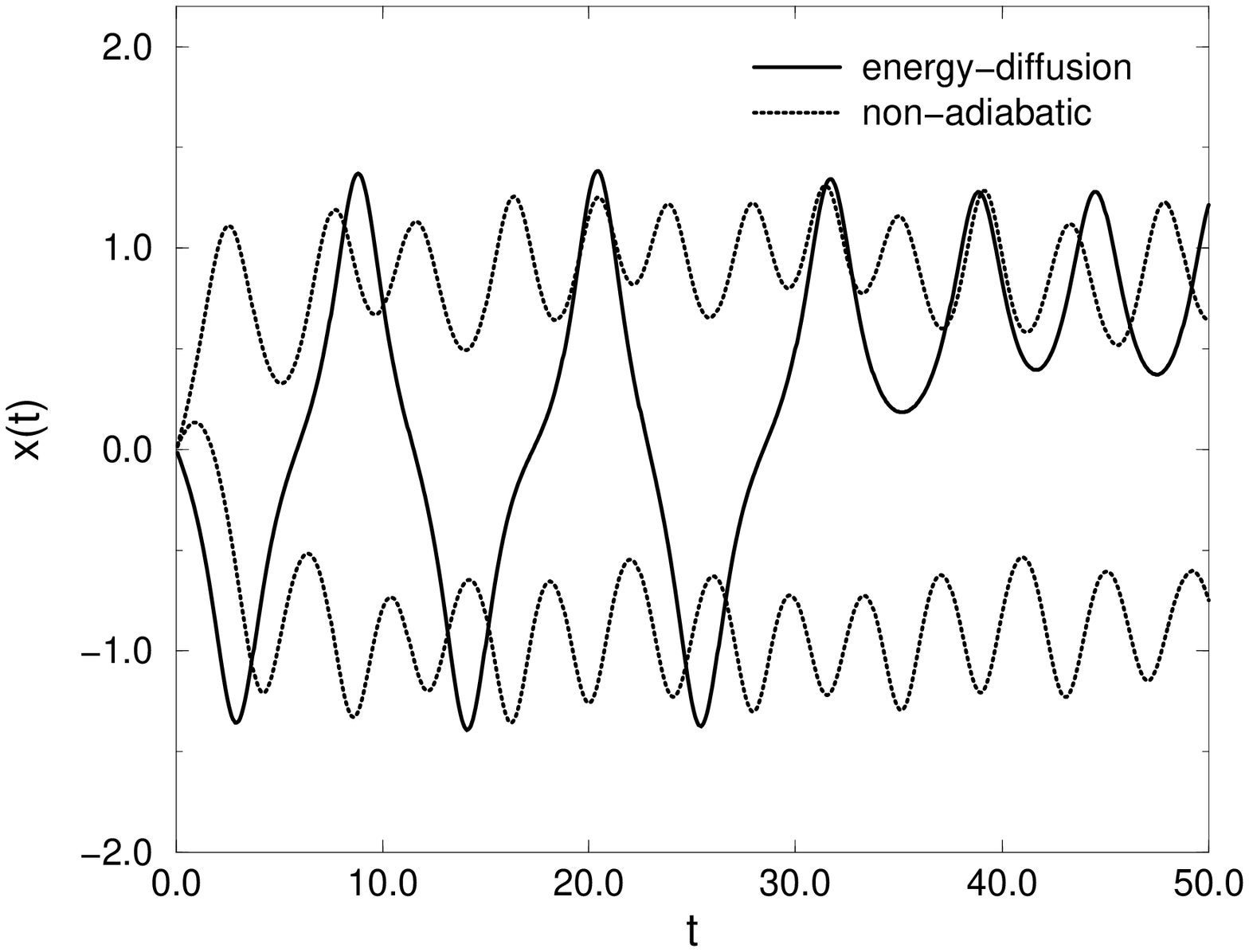}
\leavevmode
\epsfxsize = 3.0in
\epsffile{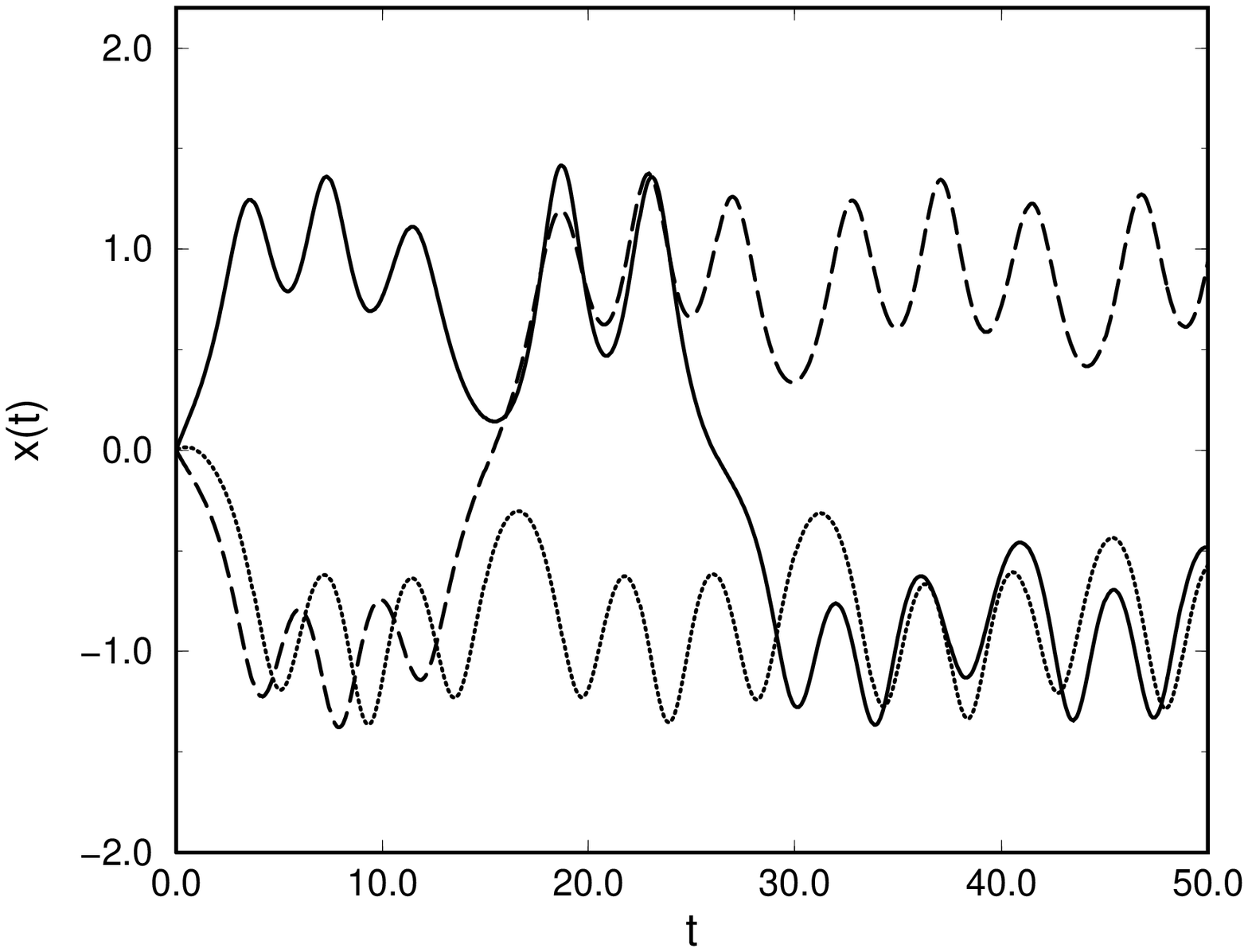}
\vspace{-0.2in}
\end{center}
\caption
{Several typical trajectories of $x$ for different regimes in the
oscillatory friction problem. The parameters are those of Fig.~\ref{ktag}.
Solid curve in left panel:
energy-diffusion-limited regime, clearly showing repeated recrossings.
Dotted curves in left panel: two examples of non-adiabatic behavior.
Right panel: several trajectories associated with the stair-like regime.}
\label{diftraj}
\end{figure}

A particularly helpful view of the process is gained by looking at
explicit trajectories, as illustrated in Fig.~\ref{diftraj}.  The
solid trajectory $x(t)$ in the left panel illustrates the typical repeated
recrossings in the low dissipation energy-diffusion-limited regime.
The dotted curves correspond to two typical trajectories in
the non-adiabatic regime.  Trajectories in this regime almost never
recross the barrier. Those that do recross
the barrier do so at short times (before straying far from $x=0$), and
typically do so only once.  These trajectories reinforce the idea that in
this regime particles are quickly trapped in one well or the other due to
the high dissipation.  As seen in the right panel of
Fig.~\ref{diftraj}, the trajectories in the stair-like regime are
considerably more complex -- this complexity distinguishes the stair-like
dynamics from the other regimes.  For example,
in this new regime one finds $x$-particles that remain localized over
one well (even though they have sufficient energy to cross the barrier)
and that after circling there several times may suddenly recross the
barrier. This behavior is not found in any other regime studied so far.

Studying additional $x$-trajectories besides those shown explicitly in
the right panel of Fig.~\ref{diftraj} leads to the realization of a number
of important points.  First, we note that
\begin{itemize}
\item the $x$-particles cross the barrier only at certain specific times.
\end{itemize}
For our typical parameters, these times
are $t=10$, $16$, $25$, $31$ and $40$ (approximately), which
coincide with the decay times $T_\kappa$
for the associated $\kappa(t)$ in Fig.~\ref{ktag}. Moreover, since
we are working with low
temperatures, the energy of the $x$-particles is typically
not large enough to recross the barrier
many times. Actually, we have observed that
\begin{itemize}
\item
most of the $x$-particles that cross the barrier
do it only once.
\end{itemize}
Indeed, only about $1\%$ of the particles show multiple
recrossings in our typical example.  The fact that most particles
do not return to their original well once they have
crossed the barrier leads to
\begin{itemize}
\item
steps rather than oscillations.
\end{itemize}
On the other hand, small friction leads to very slow energy loss,
and it is for this reason that
\begin{itemize}
\item
even at long times crossing the barrier is still possible.
\end{itemize}
This combination of features characteristic of
the energy-diffusion-limited and non-adiabatic regimes ultimately leads
to the
\begin{itemize}
\item
appearance of successive steps and plateaus.
\end{itemize}

At this point there are
two obvious questions about this regime: i) Why do we see decays
only at fairly sharply defined specific times and what are these times?
In other words, how is the period $T_\kappa$ determined?
ii) What determines the depth of each decay?
The answers to these and other questions are given in
the following subsections by considering the effects of varying the
parameters of our model.

However, before going ahead, we should first understand how
an $x$-particle can be trapped in one well in spite of
having enough energy to cross the barrier, as well as how an
$x$-particle can cross the barrier seemingly
without having enough energy to do it.
The reason for both strange situations is that the behavior of the
$x$-particles may depend strongly on that of their associated
$y$-oscillators. A given $x$-particle with energy greater than the
barrier height can be ``trapped" in one well because each time it
goes toward the barrier its coupled oscillator pulls it back.
Conversely, a given $x$-particle with apparently insufficient
energy can cross the barrier by being pulled
by its $y$ oscillator. Thus, the coupling between $x$ and $y$
may determine the times at which
the particles cross the barrier and therefore the times for the
decays of $\kappa(t)$.

\subsection{Dependence on $k$}
\label{depk}

To deduce the role of the coupling constant $k$, we depart from our typical
value ($k=0.3$) and consider the trajectories for two cases on either side
of this value but that still essentially preserve the stair-like
behavior.  We call them
the large-$k$ case ($k=0.5$) and the small-$k$ case ($k=0.1$).
The parameter $k$ determines the extent to which $x$ and $y$ particle
dynamics are in synchrony.

\begin{figure}[htb]
\begin{center}
\leavevmode
\epsfxsize = 3.0in
\epsffile{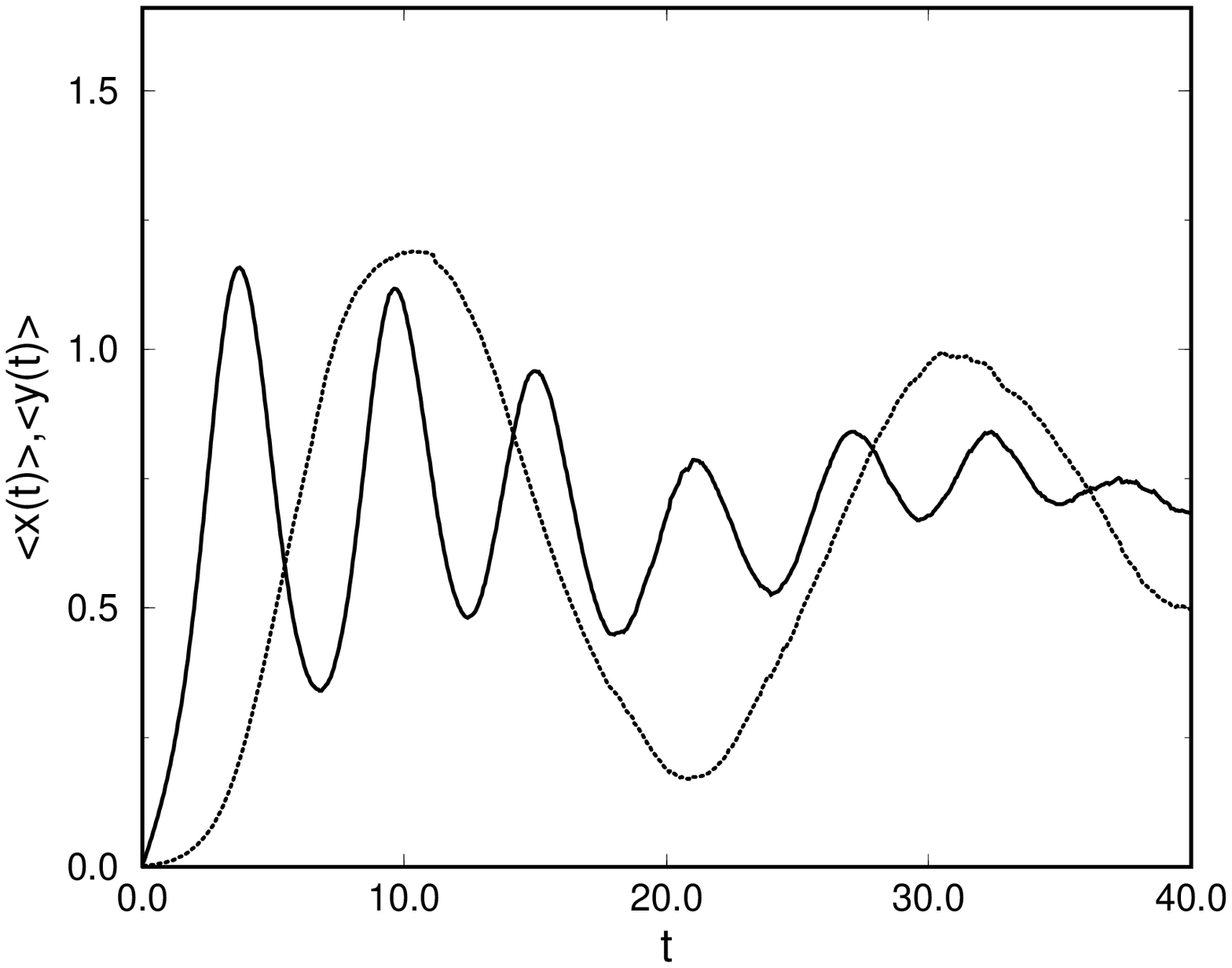}
\leavevmode
\epsfxsize = 3.0in
\epsffile{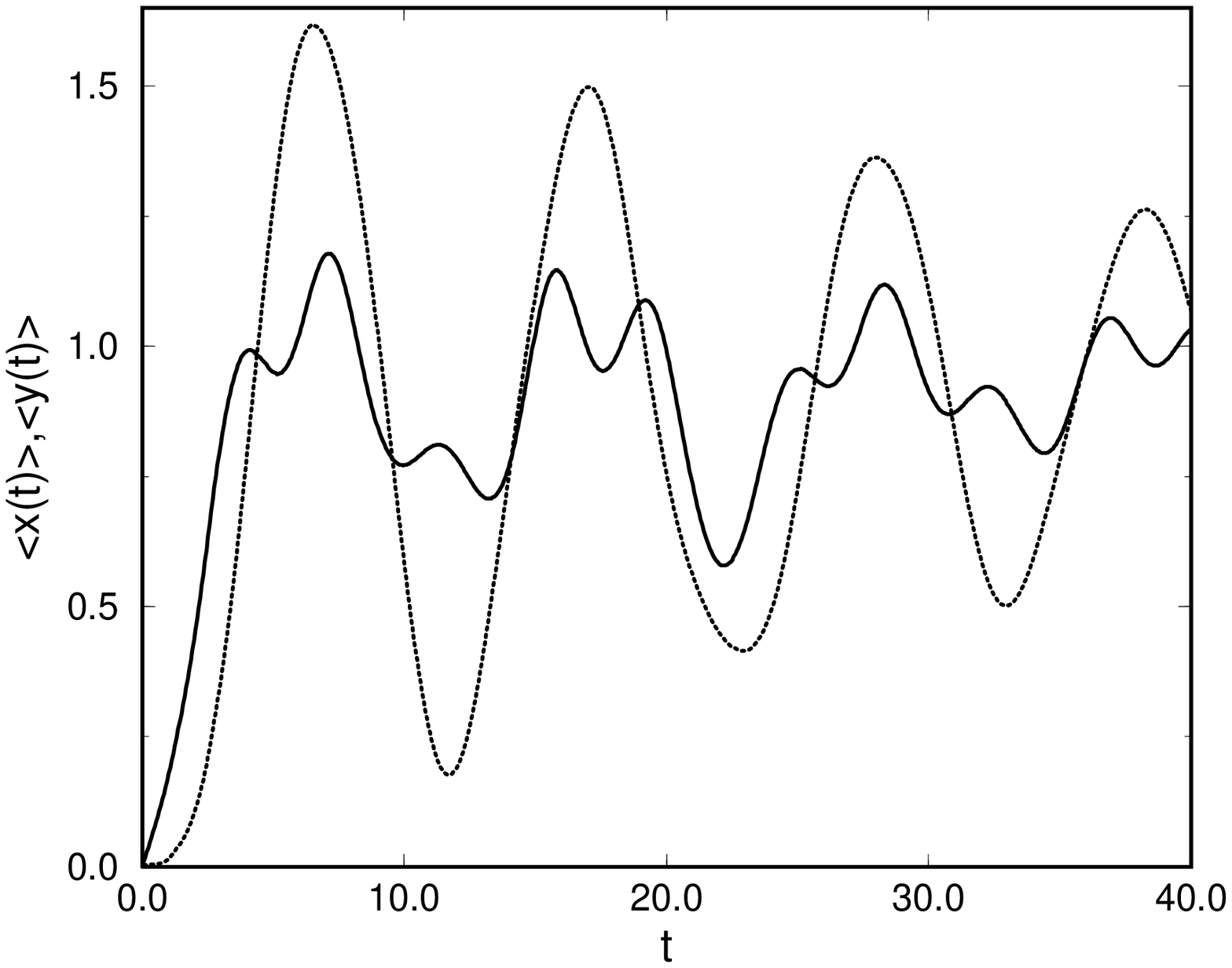}
\vspace{-0.2in}
\end{center}
\caption
{Left panel: $\left<x(t)\right>$ (solid curve)
and $\left<y(t)\right>$ (dotted curve) for the small-$k$ case.
Right panel: $\left<x(t)\right>$ (solid curve)
and $\left<y(t)\right>$ (dotted curve) for the large-$k$ case.}
\label{trajsmallk}
\end{figure}

Since $\kappa(t)$ contains information averaged over ensembles of
particles, we are interested in the average trajectories of
both $x$ and $y$ coordinates. We thus plot
$\left<x(t)\right>$
and $\left<y(t)\right>$, where $\left<\cdots\right>$ here
means an average over all the particles that are in the right well
($x>0$) at each given time.  Fig.~\ref{trajsmallk}
shows averaged trajectories for the
small-$k$ (left panel) and large-$k$ (right panel) cases.

For the small-$k$ case we readily observe that the motion of
$x(t)$ and $y(t)$ are essentially uncorrelated. Each dynamics
proceeds with a different principal frequency of oscillation. A
Fourier analysis of the trajectories reveals that the peak
frequency for $x(t)$ is $1.022$ while that of $y(t)$ is $0.306$.
These frequencies can be associated with two characteristic
frequencies of our problem. That of $x(t)$ corresponds to the
frequency of the particle in the bistable potential, namely $1.022
\approx 2\pi/T_{semi}$, where $T_{semi}$ is roughly the average
semiorbit time for an ensemble of particles above the barrier in
the double-well potential.  In Ref.~\cite{pendent1} we have shown
that the semiorbit time for a particle at an energy $\varepsilon$
above the barrier is $t_\varepsilon=\ln (16/\varepsilon)
+O(\varepsilon\ln\varepsilon)$.  An average of this time over a
thermal distribution then directly yields $T_{semi}\approx
3.35\ldots-\ln k_BT$. The frequency of $y(t)$, on the other hand,
coincides with the frequency of $\Gamma(t)$, namely $0.306 \approx
\Omega  \approx \sqrt{k+\omega^2}$. This frequency characterizes
the motion of $y(t)$ in the second equation for the extended
system, Eq.~(\ref{eqext}), when the coupling contribution is
neglected [see also Eq.~(\ref{detext})].  The $x$-particles can
cross the barrier with frequency $2\pi/T_{semi}$ since they do not
care where their associated $y$-particles are. Thus in this weakly
coupled regime
\begin{equation}
T_{\kappa} \approx T_{semi} \approx 3.35\ldots-\ln k_BT.
\label{weakly}
\end{equation}

The large-$k$ case exhibits a different behavior. In this case
$x(t)$ and $y(t)$ are essentially synchronized. We see in
Fig.~\ref{trajsmallk} that $x(t)$ has two characteristic periods:
a shorter one associated with motion in the bistable potential
($2\pi/T_{semi}$) and a longer one that matches that of $y(t)$.
The frequency of the latter is $0.624$ and coincides with
$\sqrt{k+\omega^2}\approx\sqrt{k}$. Thus, the motion of $x$ is now
dominated by the dynamics of its coupled oscillator. The
consequence of the strong coupling is that now
\begin{equation}
T_{\kappa} \approx \frac{2\pi}{\Omega} \approx \frac{2\pi}{\sqrt{k}}.
\label{strongly}
\end{equation}

\begin{figure}[!htp]
\begin{center}
\leavevmode
\epsfxsize = 3.0in
\epsffile{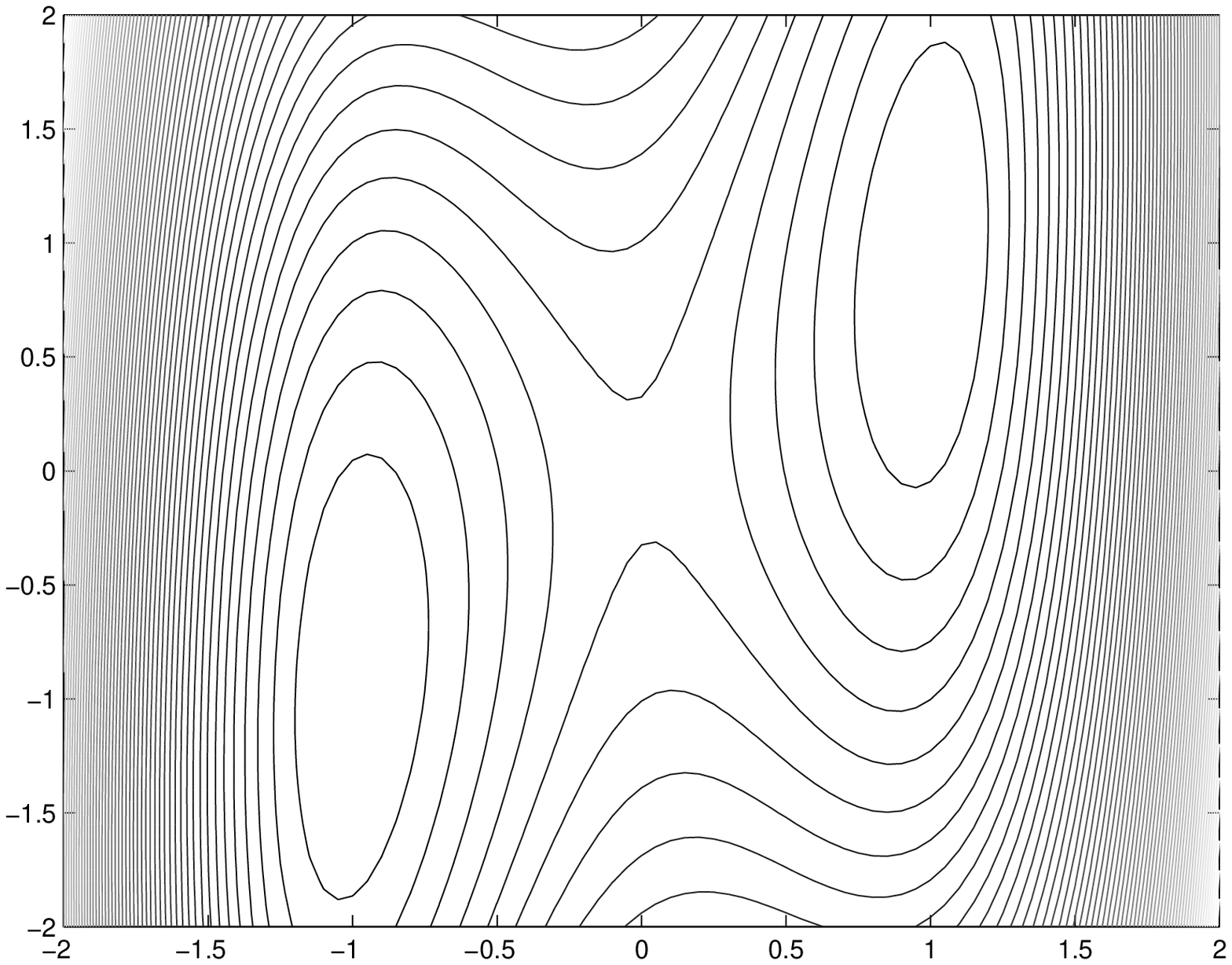}
\leavevmode
\epsfxsize = 3.0in
\epsffile{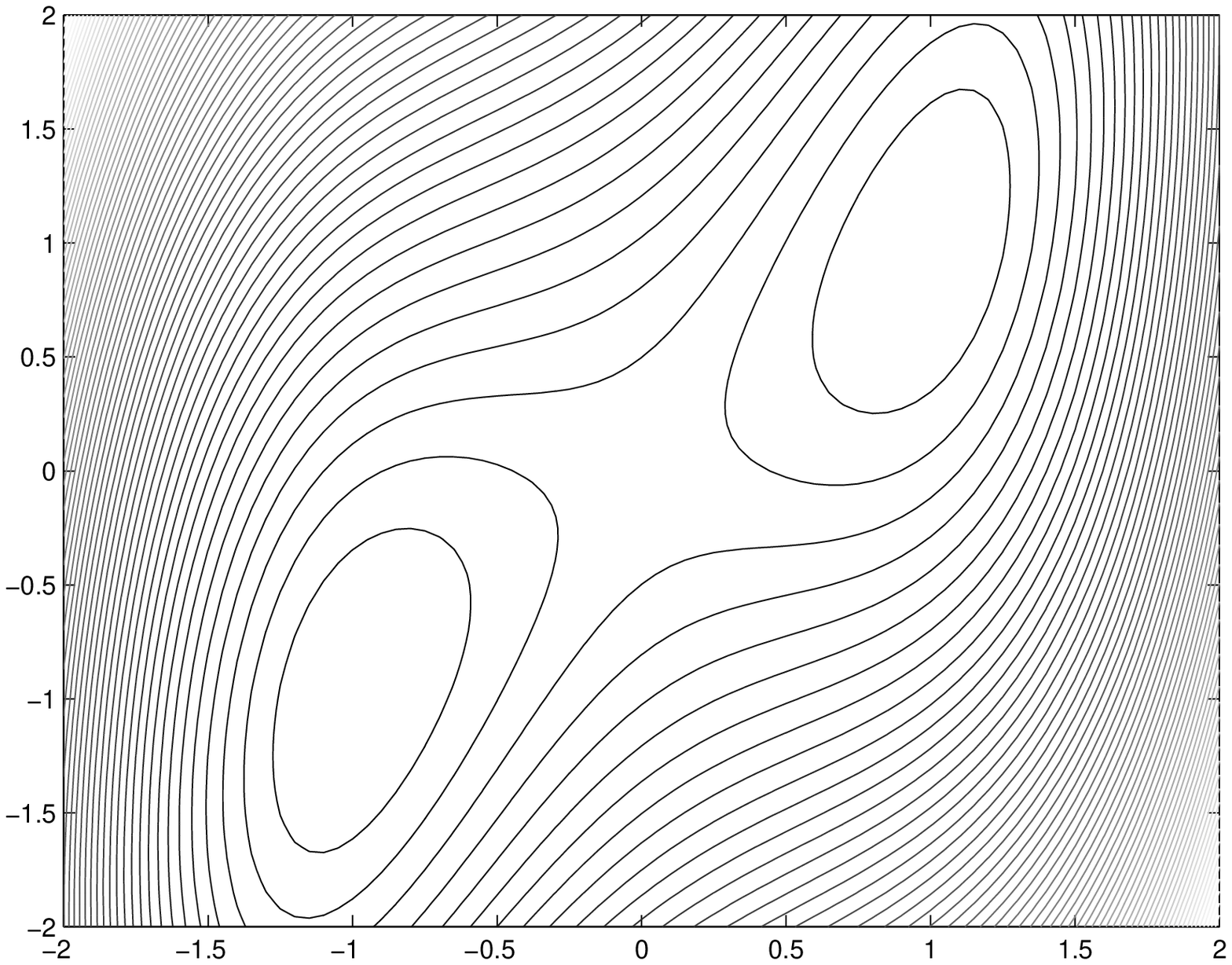}
\vspace{-0.2in}
\end{center}
\caption
{Surface contours for the potential $V(x,y)$ for the small-$k$ (left
panel, $k=0.1$) and large-$k$ (right panel, $k=0.5$) cases.  The
coordinates $x$ and $y$ are represented along the horizontal and
vertical axes respectively.}
\label{contourzb}
\end{figure}

These ideas can be further supported by considering the
two-variable potential, Eq.~(\ref{potential}), drawn in contour
form in Fig.~\ref{contourzb} for the small-$k$ (left panel) and
large-$k$ (right panel) cases.  These plots clearly illustrate the
correlations between $x$ and $y$ (or the lack thereof). When the
system is in one of the two two-dimensional wells, $x$ and $y$
remain more tightly bound in the large-$k$ case than in the
small-$k$ case. In particular, when $k$ is small the $y$-particle
can move away from $x$ even when the system has already fallen
into one well.

Further, consider the likely pathways followed by the system as it
crosses, say, from the right to the left. In the small-$k$ case
the likely path is for $y$ to decrease first (perhaps even to
negative values), followed by a change of $x$ from $x>0$ to $x<0$.
On the other hand, in the large-$k$ case it is easier for $x$ to
first move from $x>0$ to $x<0$ to be followed by $y$.  Therefore,
in the large-$k$ case the crossing rate is determined by the
frequency of $y(t)$ so that $T_\kappa\approx 2\pi/\sqrt{k}$; in
the small-$k$ case the crossing rate is limited by the motion of
$x(t)$ and hence $T_\kappa\approx T_{semi}$.

\begin{figure}[!htp]
\begin{center}
\leavevmode
\epsfxsize = 3.0in
\epsffile{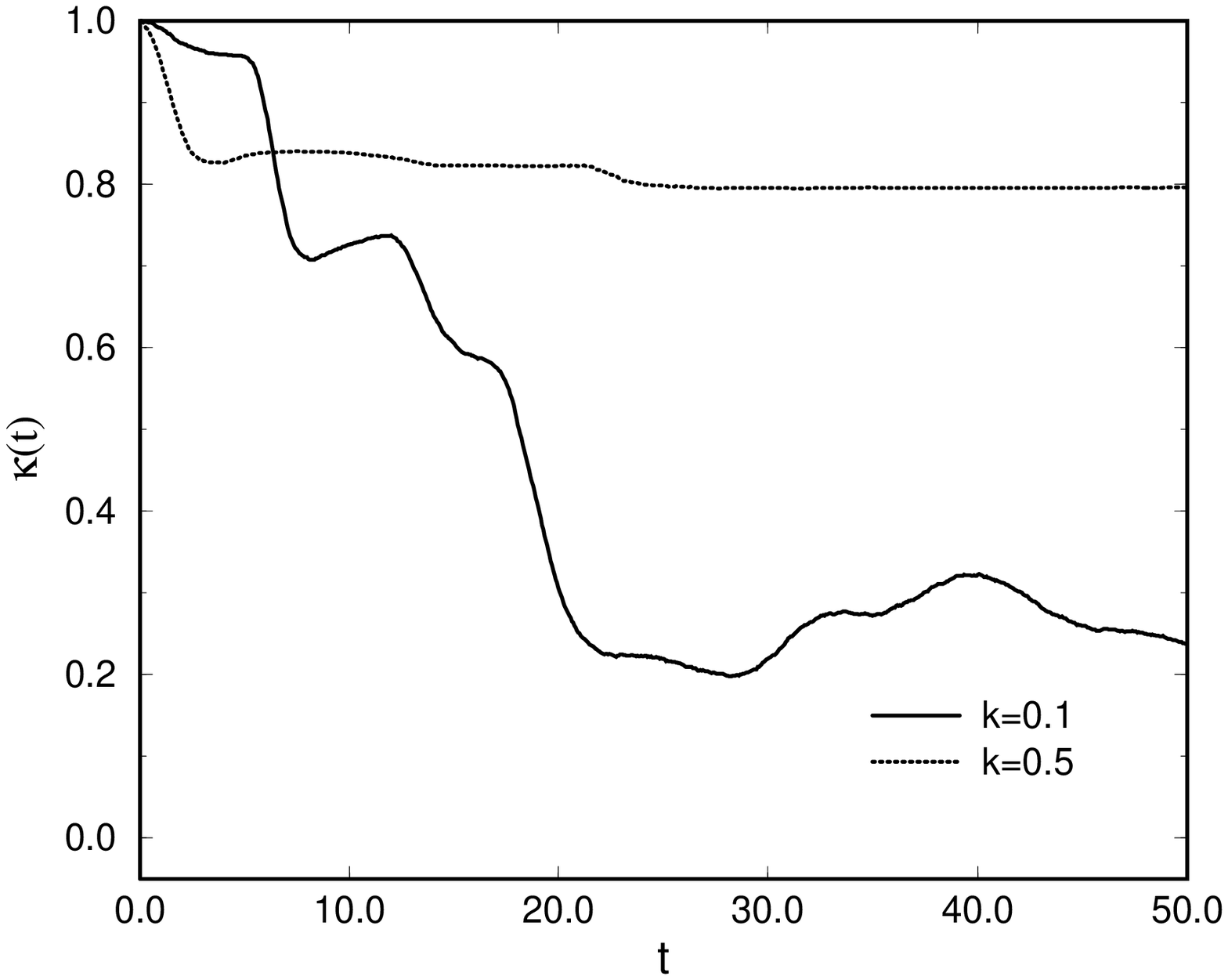}
\leavevmode
\epsfxsize = 3.0in
\epsffile{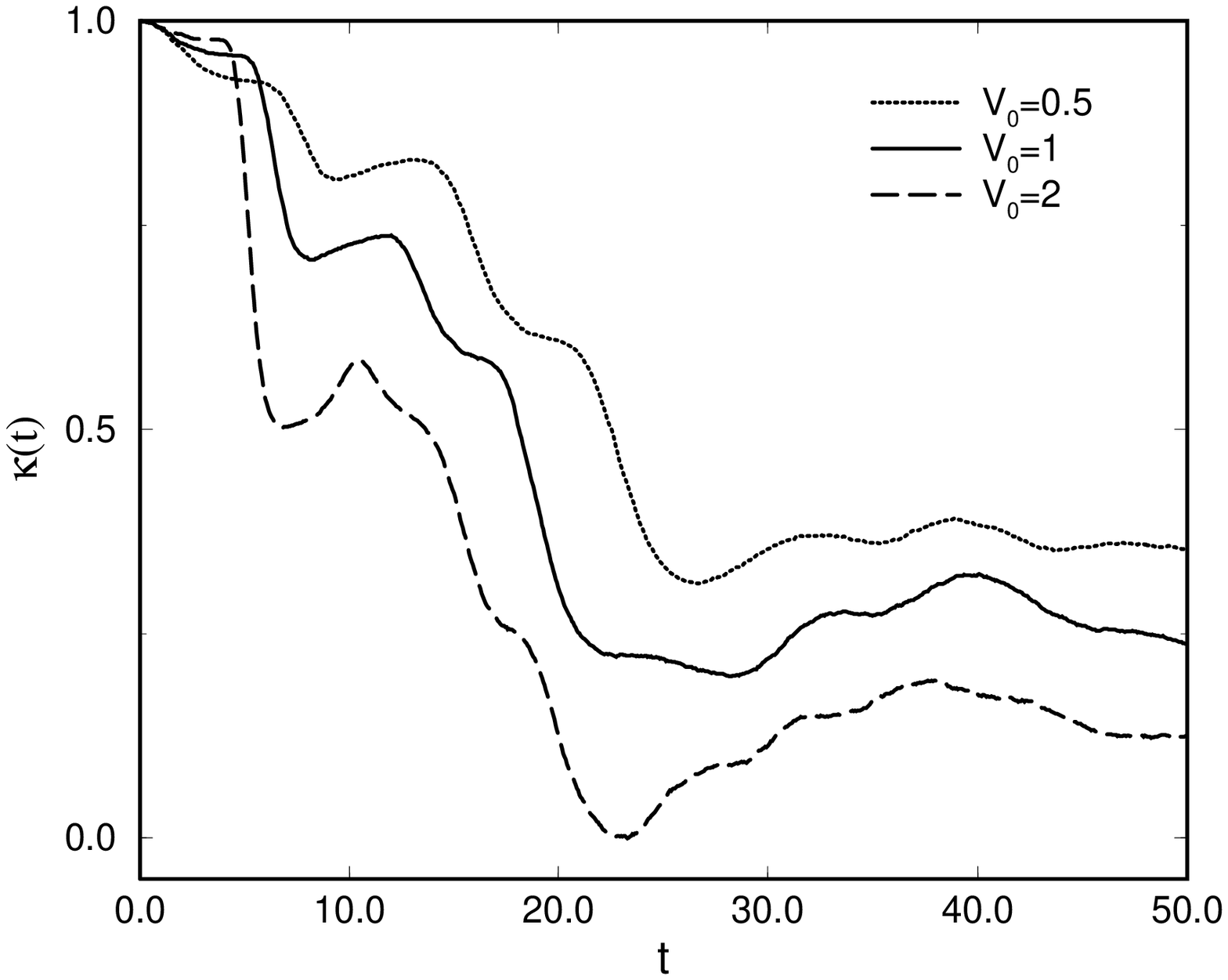}
\vspace{-0.2in}
\end{center}
\caption
{Left panel: $\kappa(t)$ for the small-$k$ and large-$k$ cases
discussed in the text. Right panel: 
Transmission coefficient for the small-$k$ case for three different
barrier heights.  The solid curve is the same as the small-$k$ case in the
left panel.}
\label{difk}
\end{figure}

Figure~\ref{difk} shows the time-dependent transmission
coefficient for the two cases.  The times $T_\kappa\approx 7.55$
(small-$k$ case) and $T_\kappa\approx 10$ (large-$k$ case)
obtained from the above arguments are consistent with the steps in
the figure (measured from mid-point to mid-point), most clearly in
the length of the first step.

In addition to the step period differences, the left panel in
Fig.~\ref{difk} 
illustrates the $k$-dependence of the depths of the steps in the stair-like
transmission coefficient.  The steps are clearly deeper when $k$ is small.
The reason for this is that the total system
loses energy by dissipation only through the $y$ coordinate.
Although both cases in the figure correspond to the same value of
$\gamma$, the $x$-particles can
retain energy for a longer time when $k$
is small.  This allows more $x$-particles to cross the barrier, and it
allows them to do so at later times.  The deeper and more numerous
clear steps in the small-$k$ case are a direct manifestation of
these features.  

Two further points should be noted.  One is the symmetry of the semiorbit 
time $t_\varepsilon$ with respect to $\varepsilon$.  That is, the semiorbit
time of a particle with an energy $\varepsilon$ {\em above} the barrier is
the same as the orbit time of a particle with energy $\varepsilon$ {\em
below} the barrier (for small $\varepsilon$).  This symmetry is important
because it allows particles to remain in synchrony; otherwise the steps in
$\kappa(t)$ would be blurred.  The other point is the dependence of
$\kappa(t)$ and consequently of the period $T_\kappa$ on barrier height.
In general, $t_\varepsilon\approx V_0^{-1/2} \ln (16V_0/\varepsilon)$
(which reduces to our previous expression when $V_0=1$), and an average of
$t_\varepsilon$ over a thermal distribution of particles above the barrier
yields the generalization of Eq.~(\ref{weakly})
\begin{equation}
T_{\kappa} \approx \frac{1}{V_0^{1/2}}\left(\ln 16V_0 + 0.5772\ldots
-\ln k_BT\right).
\label{weaklyg}
\end{equation}
The right panel in Fig.~\ref{difk} shows the transmission coefficient
in the small-$k$ case
for three values of the barrier height.  The corresponding period estimates
for $T_\kappa (V_0)$ obtained from Eq.~(\ref{weaklyg}) are
$T_\kappa(0.5)=9.70$, $T_\kappa(1.0)=7.55$, and $T_\kappa(2.0)=5.83$.
These decreasing periods with increasing barrier height are clearly
consistent with the numerical results.

\subsection{Dependence on $\gamma$}
\label{depg}

\begin{figure}[!htp]
\begin{center}
\leavevmode
\epsfxsize = 3.0in
\epsffile{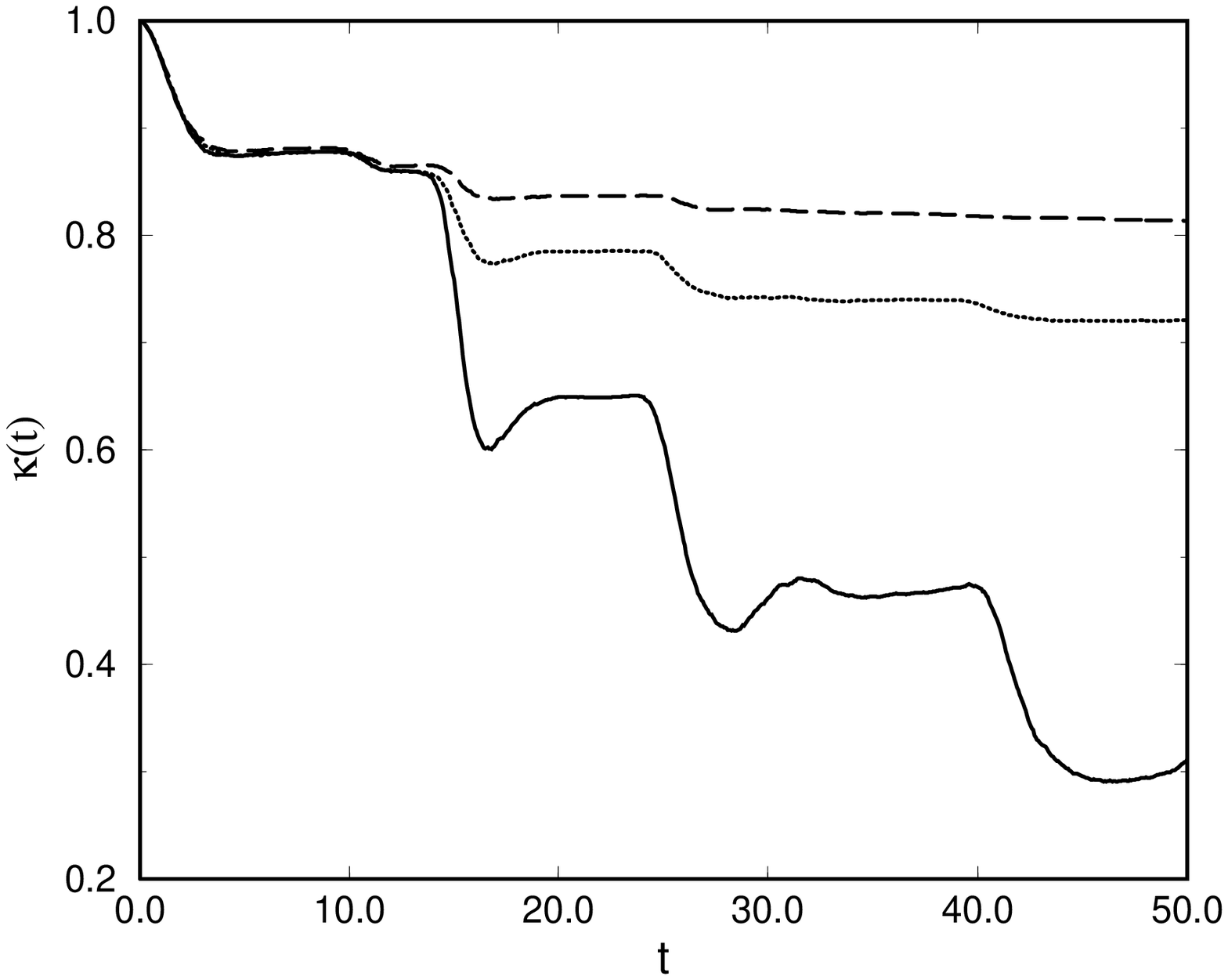}
\leavevmode
\epsfxsize = 3.0in
\epsffile{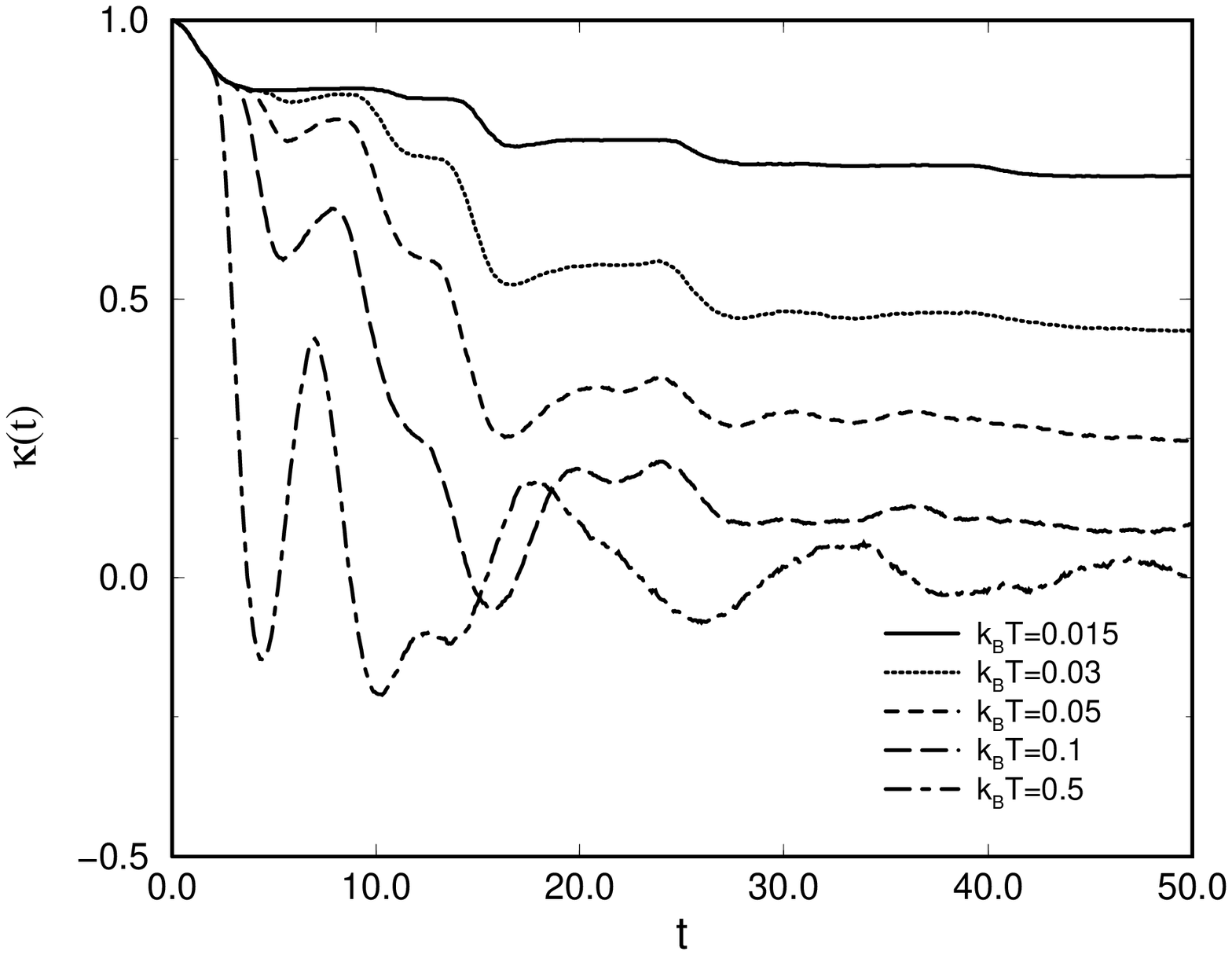}
\vspace{-0.2in}
\end{center}
\caption
{Left panel: dependence of the stair-like transmission coefficient on
the dissipation parameter $\gamma$.  The parameters are otherwise
those of our typical example
in Fig.~\ref{ktag}: $\omega^2=0.01$, $k=0.3$, $k_BT=0.015$.  Dotted curve:
$\gamma=0.05$; solid curve: $\gamma=0.01$; dashed curve: $\gamma=0.1$.
Right panel: dependence of the stair-like transmission coefficient on
temperature; the other parameters are those of
Fig.~\ref{ktag}.
}

\label{difgam}
\end{figure}

We have seen that successive steps in the transmission coefficient
arise because the particles lose their energy slowly.  This requires 
$\gamma$ to be small  -- but not too small (cf. below).  Indeed,
if $\gamma$ is decreased we expect particles to lose their energy even more
slowly, which leads to a larger number of deeper steps.  However, as
$\gamma$ continues to decrease we expect to begin to see particles that
cross the barrier more than once before becoming trapped.  This leads to
oscillations in $\kappa(t)$ and, eventually, to energy-diffusion-limited
behavior.  Deeper steps and the first appearance of small oscillations
with decreasing $\gamma$ are clearly evident in the left panel of
Fig.\ref{difgam}.

Conversely, if $\gamma$ is increased,  particles lose their energy
more rapidly, and fewer particles
cross the barrier at all; those that do so cross at most once.
In this case, as seen in Fig.~\ref{difgam},
the steps are less deep and almost disappear at long times.  Indeed,
the limit of the stair-like behavior with increasing $\gamma$ is
the monotonic non-adiabatic regime, where only a few particles
cross the barrier and they do so at very short times.

\subsection{Dependence on $k_BT$}
\label{depT}

Finally, we consider the temperature dependence of the transmission
coefficient in the stair-like regime.  As temperature is increased, all
else remaining the same, there is a greater number of more energetic
$x$-particles above the barrier.  Two aspects of their behavior dominate the
resulting transmission coefficient. One is that the particles now have
a greater range of semiorbit times $t_\varepsilon$; the other, more
important, effect is that particles are now sufficiently energetic that
they can recross the barrier more than once.  These are precisely the
features that lead to the typical oscillatory behavior of the transmission
coefficient in the energy-diffusion-limited regime, and it is towards this
behavior that the stair-like regime tends with increasing temperature.  
The right panel in Fig.~\ref{difgam} shows this progression very clearly:
the highest temperature results look very much like the earlier curves for
the energy-diffusion-limited case.  We should note the 
deeper first decay in the $k_BT=0.5$ curve than observed in our earlier
illustrations.  This is due to the fact that here we have chosen 
$\omega^2$ to be very small (a requirement for the stair-like regime).
This causes the initial thermal distribution of $y(0)$ to pull back 
$x$-particles more effectively than in our earlier example, and this in
turn leads to the deeper decay.

\subsection{Arguments in the Reduced System}
\label{argred}

Since the extended [Eq.~(\ref{eqext})] and reduced [Eq.~(\ref{generic1})]
systems are entirely equivalent, it is of course possible to explain the
new stair-like regime in terms of quantities and equations associated with
the reduced representation.  It is perhaps somewhat more cumbersome and
less transparent, but the extended representation analysis offers a helpful
guide.  For example, it is useful to realize that a ``negative friction"
[i.e., a negative value of the memory kernel in Eq.~(\ref{generic1})] in
the reduced representation is associated with the situation where in the
extended system the $y$-coordinate pulls the $x$-particle in the direction
of $\dot{x}$.

To explain the different decay periods $T_\kappa$ in the reduces
representation we note that the memory kernel $\Gamma(t)$ is proportional
to $k^2$ and the random force $F(t)$ is proportional to $k$.  To cross the
barrier, an $x$-particle must be moving towards $x=0$.  When $k$ is small,
the dynamics of $x$ as it moves in the barrier region is thus dominated by
the bistable potential $V_{eff}(x)$.  In particular, the decay periods in
$\kappa(t)$ are determined by the frequency of the particles moving in the
bistable potential with energies slightly larger and slightly smaller than
the barrier height, which directly leads again to the earlier estimate
$T_\kappa = T_{semi}$ [cf. Eq.~(\ref{weakly})].  The slow dissipation of
$x$-energy associated with small $k$ allows for many deep steps in
$\kappa(t)$.  

As $k$ increases, the bistable potential becomes relatively less important
and the first and third terms on the right of Eq.~(\ref{generic1})
increasingly dominate the dynamics of $x(t)$.  The steps in $\kappa(t)$
then acquire the period $T_\kappa \approx 2\pi/\Omega \approx
2\pi/\sqrt{k}$ associated with the friction kernel.  The more rapid
dissipation of $x$-energy associated with larger $k$ leads to a small
number of shallow steps.

The dependence on $\gamma$ in this representations is quite clear.  When
$\gamma$ increases, the memory kernel decays more rapidly and the
oscillations in $\Gamma(t)$ become irrelevant, thus leading to
non-adiabatic behavior of $\kappa(t)$.  Decreasing $\gamma$, on the other
hand, leads to pronounced oscillations (and at times negative values) of
$\Gamma(t)$.  As a result, even particles that start out with energies too
low to cross the barrier early may do so at a later time, thus explaining
the step structure of $\kappa(t)$.  The temperature dependence can also be
understood: for a given (low) $\gamma$, increasing the temperature leads to
a greater number of particles above the barrier that can recross more than
once before becoming trapped.  The steps then become oscillations and the
energy-diffusion-limited behavior is recovered.

\section{Conclusions}
\label{concl}

In this work we have analyzed the time dependent transmission coefficient
for the capture of a particle in one or the other well of a bistable
potential as described by
the generalized Kramers equation Eq.~(\ref{generic1}) with an
oscillatory memory kernel.   The time dependence of the transmission
coefficient depends sensitively on the parameters of the model.  The
equivalence of this model to an ``extended system" wherein the reaction
coordinate is linearly coupled to a nonreactive coordinate which is in turn
coupled to a heat bath, Eq.~(\ref{eqext}), facilitates the understanding
of the various time dependences that are observed.  

The different behaviors observed for the transmission coefficient in
various parameter regimes are summarized in Fig.~\ref{ktag}.  The
non-adiabatic (monotonic decay) and energy-diffusion-limited (oscillatory
decay) behaviors have been encountered earlier in the classic Kramers
problem~\cite{Montg,Borgis,pendent1,Kohen} and in the generalized Kramers
problem with an exponential
memory kernel~\cite{Montg,Borgis,pendent2,Kohen}.  The non-adiabatic decay
is observed when the reaction coordinate loses its energy rapidly, so that
particles cross the barrier only at early times and
at most once before becoming trapped.  The
oscillatory behavior is observed when the reaction coordinate loses its
energy slowly, thus allowing several recrossings of the barrier before
trapping.  These regimes are associated with parameter values that suppress
the oscillations of the memory kernel.  There is a third behavior observed
with exponential friction, the caging regime, which is not observed with an
oscillatory memory friction.  

The third behavior shown in Fig.~\ref{ktag}, which consists of a stair-like
decay of the transmission coefficient, is peculiar to the oscillatory
memory friction and occurs when the oscillations in the memory kernel are
pronounced.  This behavior is observed when particles cross the barrier at
most once, but not necessarily at early times.  In turn, this can be
explained by the fact that particles that at one time may not have enough
energy to cross the barrier may acquire sufficient energy to do so later via
their coupling to the nonreactive coordinate (or, equivalently, when the
oscillations in the memory kernel periodically lead to negative values
of the kernel).  Although the particles cross the barrier at most once, and
not necessarily at early times, the crossing events can only occur at
fairly sharply defined time intervals that we call $T_\kappa$.  Hence the
appearance of fairly sharp steps in the transmission coefficient.
We explain in detail the conditions that lead to the
stair-like behavior, the way in which the step time $T_\kappa$ and the
step depths depend on the parameters of the system, and the way in which
this behavior tends to the energy-diffusion-limited or non-adiabatic
cases as parameters are modified.

If there were no barrier crossings at all in the Kramers problem,
the transmission coefficient would be unity.  Single barrier crossings
only at
early times lead to monotonic decay of the transmission coefficient.
Single recrossings that are possible only at specified time intervals
$T_\kappa$ lead to the new stair-like regime.  Multiple recrossings
lead to oscillatory behavior.  The time dependence of the transmission
coefficient clearly provides an interesting mirror for the barrier
crossing dynamics of the generalized Kramers problem.

\section*{Acknowledgments}
One of us (R. R.) gratefully acknowledges the support of this
research by the Ministerio de Educaci\'{o}n y Cultura through 
Postdoctoral Grant No. PF-98-46573147.
This work was supported in part by the U. S. Department of Energy under
Grant No. DE-FG03-86ER13606, and in part by the Comisi\'on
Interministerial de Ciencia y Tecnolog\'{\i}a (Spain)
Project No. DGICYT PB96-0241.

\appendix
\setcounter{equation}{0}
\renewcommand{\theequation}{\thesection\arabic{equation}}

\section{Extended vs Reduced Model}
\label{appa}

Here we present the analytical details connecting Eqs.~(\ref{generic1})
and (\ref{eqext}).  Formal solution of Eq.~(\ref{eqext}) gives
\begin{align}
y(t)& = \frac{v_{y\circ} - y_\circ \lambda_2}{\lambda_1 - \lambda_2}\,
e^{\lambda_1 t} + \frac{y_\circ \lambda_1 - v_{y\circ}}{\lambda_1 -
\lambda_2}\, e^{\lambda_2 t}
\notag \\ \notag \\
&+
\frac{1}{\lambda_1 - \lambda_2} \int_0^t dt' \left( e^{\lambda_1
(t-t')} - e^{\lambda_2 (t-t')} \right) [ k x(t') + f(t') ] ,
\end{align}
where the first two terms on the right hand side
correspond to the homogeneous solution that
depends on the initial conditions $y(0)\equiv y_\circ$
and $\dot{y}(0)\equiv v_{y\circ}$. The last two
terms correspond to the inhomogeneous solution. The term proportional
to $k$ leads to the memory friction term and the term containing
$f(t')$ is associated with the colored noise in the reduced model.
The roots $\lambda_i$ are
\begin{equation}
\lambda_{1,2} = - \frac{\gamma}{2} \pm
\sqrt{\left( \frac{\gamma}{2} \right) ^2 - \omega^2 - k}.
\label{l1}
\end{equation}

Substitution of this formal solution in Eq. (\ref{eqext}) and
regrouping of terms directly leads to the
reduced model (\ref{generic1}) with the memory kernel
\begin{equation}
\Gamma(t) = \frac{k^2}{\lambda_2 - \lambda_1} \left( \frac{e^{\lambda_1
t}}{\lambda_1} - \frac{e^{\lambda_2 t}}{\lambda_2} \right).
\label{genfric}
\end{equation}
We also get the explicit form
for the effective potential of the reaction coordinate,
\begin{equation}
V_{eff}(x)=V(x) + \frac{1}{2} \frac{\omega^2 k}{\omega^2+k}\, x^2.
\label{efbistablea}
\end{equation}
However, the following extra initial conditions must be
fulfilled in order to avoid transient terms in the reduced model:
\begin{equation}
x(0) = 0 \, ; ~  < v_{y\circ} > = < y_\circ > = 0 \, ; ~
< v_{y\circ}^2 > = k_B T \, ; ~
< y_\circ^2 > = \frac{k_B T}{ \omega^2 + k}.
\end{equation}
The brackets here indicate averages over initial distributions.
The following initial distributions for the solvent
coordinate are consistent with these requirements:
\begin{equation}
P(y_\circ) = \sqrt{\frac{\omega^2 +k}{2 \pi k_B T}}~
\exp\left(-\frac{(\omega^2 +k)y_\circ^2}{2 k_B T}\right)
\label{y0}
\end{equation}
and
\begin{equation}
P(v_{y\circ}) = \frac{1}{\sqrt{2 \pi k_B T}} ~ \exp
\left(- \frac{v_{y\circ}^2}{2 k_B T}\right).
\label{vy0}
\end{equation}
The explicit reduction (integration) thus readily leads
to the observation that the initial conditions for $y$
are the thermalized solutions of the homogeneous
differential equation
\begin{equation}
\ddot{y} + \gamma \dot{y} + (\omega^2 +k ) y = 0.
\label{homy}
\end{equation}
That is, we must thermalize the solvent coordinate evolving in
the combined intrinsic and coupling potential with
the reaction coordinate fixed at $x=0$.

At this point, a distinction should be made between the following two
behaviors of the friction kernel.  The first is
the underdamped case, where the condition
\begin{equation}
\Omega^2\equiv \omega^2 +k -\left(\frac{\gamma}{2} \right) ^2 >0
\label{unda}
\end{equation}
leads to complex values for $\lambda_1$ and $\lambda_2$,
\begin{equation}
\lambda_{1,2} = - \frac{\gamma}{2} \pm i\Omega,
\label{lund}
\end{equation}
which in turn leads to
a trigonometric form for the memory kernel (sometimes
called the ``trigonometric case"):
\begin{equation}
\Gamma(t) = \frac{k^2}{\omega^2 +k}\,
e^{- \frac{\gamma}{2} t}\, \left( \frac{\gamma} {2 \Omega}
\sin \Omega t  + \cos \Omega t \right).
\label{trigfrica}
\end{equation}
The second behavior, the overdamped case, results when
\begin{equation}
\Lambda^2\equiv \left( \frac{\gamma}{2} \right) ^2 - \omega^2 - k \geq 0.
\label{over}
\end{equation}
In this case the values of $\lambda_1$ and $\lambda_2$ are real,
\begin{equation}
\lambda_{1,2} = - \frac{\gamma}{2} \pm \Lambda,
\label{lover}
\end{equation}
and therefore the memory kernel has a hyperbolic form (sometimes
called the ``hyperbolic case"):
\begin{equation}
\Gamma(t) = \frac{k^2}{\omega^2 +k}\,
e^{- \frac{\gamma}{2} t}\, \left( \frac{\gamma} {2 \Lambda}
\sinh \Lambda t  + \cosh  \Lambda t  \right).
\label{hypfric}
\end{equation}

\end{spacing}

\begin{thebibliography}{999}

\bibitem{Kramers} H.A. Kramers, Physica {\bf 7}, 284 (1940).

\bibitem{Grote} R. F. Grote and J. T. Hynes, J. Chem. Phys. {\bf 73}, 2715
(1980).

\bibitem{Hanggi}
P. H\:{a}nggi, P. Talkner and M. Borkovec, Rev. Mod.
Phys. {\bf 62}, 251 (1990).

\bibitem{Melnikov2} V. I. Melnikov, Phys. Report {\bf 209}, 1 (1991).

\bibitem{Pollak}
E. Pollak, in {\em Activated Barrier Crossing}, eds. P. H\:{a}nggi and G.
Fleming (World Scientific, Singapore, 1992).

\bibitem{Tuckerrev}
S. C. Tucker, in {\em New Trends in Kramers Reaction Rate Theory:
Understanding Chemical Reactivity}, eds. P. Talkner and P. H\:{a}nggi
(Kluwer Academic, Dordrecht, 1995).

\bibitem{Melnikov1} V. I. Melnikov and S. V. Meshkov, J. Chem. Phys. {\bf
85}, 1018 (1986).

\bibitem{Pollak1} E. Pollak, H. Grabert and P. H\:{a}nggi, J. Chem. Phys. {\bf
91}, 4073 (1989).

\bibitem{Pollak2} E. Pollak and P. Talkner, Phys. Rev. E {\bf 47}, 922
(1993).

\bibitem{Melnikov3} V. I. Melnikov, Phys. Rev. E {\bf 48}, 3271 (1993).

\bibitem{pendent1} J. M. Sancho, A. H. Romero and K. Lindenberg, J. Chem.
Phys. {\bf 109}, 9888 (1998).

\bibitem{pendent2} K. Lindenberg, A. H. Romero and J. M. Sancho, to appear
in Physica D.

\bibitem{Montg} J.A. Montgomery, Jr, D. Chandler, and B. J. Berne, J.
Chem.Phys. {\bf 70}, 4056 (1979).

\bibitem{Borgis}
D. Borgis and M. Moreau, Mol. Phys. {\bf 57}, 33 (1986).

\bibitem{Tucker}
S. Tucker, J. Chem. Phys. {\bf 101} (1994) 2006; S. K. Reese, S. Tucker,
and G. K. Schenter, J. Chem. Phys. {\bf 102} (1995) 104.

\bibitem{Kohen} D. Kohen and D.J. Tannor, J Chem. Phys. {\bf 103}, 6013
(1995).


\bibitem{Strauba} J. E. Straub and B.J. Berne, J. Chem. Phys. {\bf
83}, 1138 (1985).

\bibitem{Straubb} J. E. Straub, D.A. Hsu, and B.J. Berne, J. Phys. Chem.
{\bf 89}, 5188 (1985).

\bibitem{Gard} T.C. Gard, {\em Introduction to Stochastic Differential
Equations}, Marcel Dekker, Vol.114 of {\em Monographs and Textbooks in
Pure and Applied Mathematics} (1987).

\bibitem{Toral94} R. Toral, in {\em Computational Field Theory and Pattern
Formation}, 3rd Granada Lectures in Computational Physics, Lecture Notes in
Physics Vol. 448 (Springer Verlag, Berlin, 1995).

\end{thebibliography}
\end{document}